%% file: paper.tex
\documentclass[]{aa}

\usepackage[utf8]{inputenc}
\usepackage{graphicx}
\graphicspath{{figures/}}
\usepackage{natbib}
\usepackage{hyperref}
\hypersetup{pdfpagemode = {UseNone},
            pdftitle = {Survival of planet-induced vortices in 2D disks},
            pdfauthor = {Thomas Rometsch, Alexandros Ziampras, William B{\'e}thune, Wilhelm Kley},
            pdfsubject = {},
            pdfview = {FitH},
            pdfstartview = {FitH},
            colorlinks = {true},
            linkcolor = [rgb]{0,0.35,0.7},
            citecolor = [rgb]{0,0.35,0.7},
            filecolor = [rgb]{0.61,0,0},
            urlcolor = [rgb]{0.61,0,0},
           }
\usepackage{txfonts}
\usepackage{enumitem}
\usepackage{placeins}
\usepackage{longtable}

\usepackage{bm} %bold math
\usepackage{upgreek} %for stress tensor

\begin{document}

\title{Survival of planet-induced vortices in 2D disks}

\author{Thomas Rometsch
	\and
	Alexandros Ziampras
	\and
	Wilhelm Kley
	\and
	William B{\'e}thune
}

\institute{Institut für Astronomie und Astrophysik, Universit\"at T\"ubingen,
	Auf der Morgenstelle 10, 72076 T\"ubingen, Germany\\
	\email{\{thomas.rometsch, alexandros.ziampras\}@uni-tuebingen.de}
}

\date{\today}

\abstract
{% context heading (optional)
	Several observations of protoplanetary disks display non-axisymmetric features, often interpreted as vortices. Numerical modeling has repeatedly shown that gap-opening planets are capable of producing large and long-lasting vortices at their outer gap edge, making massive planets popular candidates as the source of such features.
}
{% aims heading (mandatory)
	We explore the lifetime of vortices generated by Jupiter-sized planets as a function of the thermal relaxation timescale, the level of turbulence, and the effect of disk self-gravity.
}
{% methods heading (mandatory)
	We conduct 2D numerical simulations using the hydrodynamics codes \texttt{PLUTO} and \texttt{FARGO}, scanning through several physical and numerical parameters.
	Vortex properties are automatically extracted from thousands of simulation snapshots.
}
{% results heading (mandatory)
	We find that vortices that spawn at the outer gap edge can survive for about 100--3000 planetary orbits, with the shortest lifetimes occurring for moderately efficient dissipation and cooling. However, we also observe a different regime of long-lasting vortices with lifetimes of at least 15\,000 orbits for very low viscosity and very short thermal relaxation timescales. Disk self-gravity significantly shortens the lifetime of regular vortices but still allows long-lived ones to survive.
}
{% conclusions heading (optional), leave it empty if necessary
	Our results suggest that the cooling timescale plays an important role in vortex formation and lifetime and that planet-generated vortices should be observable at large distances from the star for typical thermal relaxation timescales and low turbulence levels.
}

\maketitle
%
%________________________________________________________________

\section{Introduction}

Planets are born and grow in accretion disks around young stars.
This is supported by observations of protoplanets embedded in a disk of gas and dust captured during their growth phase \citep[e.g.][]{keppler_discovery_2018}.
A protoplanet interacts with the disk around it in every stage of its growth \citep{kley_planet-disk_2012}
for example via exchange of angular momentum. This results in the launching of spiral arms \citep{ogilvie-lubow-2002} and, if the planet is massive enough,
the opening of a gap and, in some cases, the formation of multiple rings around the planet's orbit \citep{rafikov-2002}.
The number of spirals, gaps, and rings as well as their contrast scales with the planet's mass,
such that Jupiter-sized planets can have a strong impact on their environment in the right conditions,
possibly resulting in multiple ring-like and non-axisymmetric observable features \citep{zhang-zhu-2020, miranda-rafikov-2020a}.
This makes the planet--disk interaction scenario a popular interpretation for the numerous high-fidelity ALMA observations of such features.

One promising scenario to explain observational asymmetries is the existence of vortices because they naturally accumulate dust at the pressure maxima in their center \citep[see for example][]{marel_major_2013,bae_planetary_2016,perez_disk_2018,hammer_observational_2019,barge-sommeria-1996}.

Among the various ways to form vortices, the Rossby-wave instability \citep[RWI,][]{lovelace_rossby_1999} is particularly relevant in the vicinity of gaps. The RWI readily happens in 2D disks at the outer and inner edge of planet-opened gaps \citep{li_potential_2005,val-borro_vortex_2007}. Additional mechanisms that could be relevant in this context are the subcritical baroclinic instability \citep[SBI,][]{klahr_turbulence_2003, lesur_subcritical_2010} and the zombie-vortex instability \citep[ZVI,][]{marcus_zombie_2015,marcus_zombie_2016}.
Vortices are then susceptible to viscous spreading as well as secondary instabilities such as the elliptical instability \citep{lesur_stability_2009}, which cause vortex decay.
The lifetime of vortices is therefore determined by a competition between vortex-forming and -decaying mechanisms.

Aside from possibly causing observable features in the disks, vortices can also affect planet migration in a stochastic fashion \citep{regaly_trapping_2013,ataiee_planet-vortex_2014,mcnally_migrating_2019} and even cause temporary outward migration \citep{lega-etal-2021} for otherwise inwardly-migrating planets. Understanding their formation pathways and lifetimes is therefore critical to the modeling of planet migration using global, low-viscosity simulations.

In previous numerical studies, vortex properties have been found to depend on various physical processes such as turbulent viscosity and disk self-gravity.
Lower viscosity allows vortices to live longer \citep{godon_vortices_1999,val-borro_vortex_2007,ataiee_asymmetric_2013,fu_long-term_2014,regaly_interpreting_2017} whereas the inclusion of self-gravity tends to weaken vortices, shortening their lifespan \citep{lin_effect_2011, zhu_gas_2016,regaly_vortex_2017,pierens_evolution_2018}.

In recent numerical studies, radiative effects have been discovered to have a significant impact on the gap-opening capabilities of planets and therefore the structure of said gaps \citep{ziampras-etal-2020b,miranda-rafikov-2020b}, affecting the development of the RWI and by extension vortices around their edge \citep{tarczay-nehez_vortex_2020}.
The present study aims to investigate the role of radiative effects for properties of vortices created by planets.
More precisely, we explore how the thermal relaxation timescale of the gas affects the lifetime of vortices created during the growth of Jupiter-sized planets.

The impact of thermal relaxation on vortex formation and lifetime was studied for nearly inviscid disks by \cite{les_gap_2015}, 
and recently \citet{fung_cooling-induced_2021} ran 2D shearing box simulations of RWI-induced vortices. 
As their simulations did not include a planet, the RWI was triggered by an artificial density bump.
They described a baroclinic effect that spins down vortices where the decay is fastest for thermal relaxation times of the order of a tenth of the vortex turnover time.

We ran a suite of global two-dimensional hydrodynamics simulations with an embedded Jupiter-sized planet, which naturally creates vortices in the disk, for different choices for the turbulent viscosity and the thermal relaxation timescale, among other physical parameters.
The results of these simulations are then post-processed with our newly-developed pipeline for the detection and characterization of vortices.

In Sect.~\ref{sec:physics-numerics} we describe our physical model and numerical setup. We present a typical life track of a vortex in our models in Sect.~\ref{sec:lifetrack}, report the dependence of vortex properties on physical parameters in  Sect.~\ref{sec:results}, and present the case of long-lived vortices in Sect. \ref{sec:long-lived}. We discuss and comment on our findings in Sect.~\ref{sec:discussion}. Finally, Sect.~\ref{sec:summary} contains a summary of our main results and our conclusions.

\section{Physics and numerics}
\label{sec:physics-numerics}

In this section, we describe the physical and numerical framework that we used in our simulations. We justify the approximations in our model, explain in detail the initialization process, and list technical parameters such as our grid setup and parameter space.

\subsection{Hydrodynamics}
\label{sec:hydrodynamics}

We consider a thin disk of neutral, ideal gas with adiabatic index $\gamma=7/5$ and mean molecular weight $\mu=2.353$ that is orbiting around a star with one solar mass $M_\star=\mathrm{M}_\odot$. The two-dimensional, vertically integrated Navier-Stokes equations in a polar coordinate system $\{r, \phi\}$ read
\begin{subequations}
	\label{eq:navier-stokes}
	\begin{align}
		\label{eq:navier-stokes-continuity}
		\frac{\partial\Sigma}{\partial t} + \vec{u}\cdot\nabla\Sigma & = -\Sigma \nabla\cdot\vec{u}
	\end{align}
	\begin{align}
		\label{eq:navier-stokes-momentum}
		\Sigma\frac{\partial\vec{u}}{\partial t} + \Sigma(\vec{u}\cdot\nabla)\vec{u} & = -\nabla p + \Sigma \vec{g} + \nabla \cdot \bm{\upsigma}
	\end{align}
	\begin{align}
		\label{eq:navier-stokes-energy}
		\frac{\partial (\Sigma\varepsilon)}{\partial t} +\vec{u}\cdot\nabla(\Sigma\varepsilon) & = -\gamma \Sigma\varepsilon \nabla\cdot\vec{u} + Q_\mathrm{visc} + Q_\mathrm{relax}
	\end{align}
\end{subequations}
where $\vec{u}=(u_r, u_\phi)$ and $\varepsilon$ are the velocity and specific internal energy of the gas evaluated at the midplane, and $\Sigma$ is the surface density.
The vertically integrated pressure $p$ is defined through the ideal gas law $p = (\gamma-1)\Sigma\varepsilon = \mathrm{R}_\mathrm{g}\Sigma T/\mu$, with $\mathrm{R}_\mathrm{g}$ being the gas constant and $T$ the gas temperature. 
The isothermal sound speed of the gas is then given by $c_\mathrm{s,iso} = \sqrt{p/\Sigma} = \sqrt{\mathrm{R}_\mathrm{g} T/\mu}$ and relates to the adiabatic sound speed $c_\mathrm{s}$ as $c_\mathrm{s,iso} = c_\mathrm{s}/\sqrt{\gamma}$. 
For a disk in Keplerian motion and vertical hydrostatic equilibrium, we can also write $c_\mathrm{s,iso} = H\Omega_\mathrm{K}$, where $\Omega_\mathrm{K} = \sqrt{\mathrm{G}M_\star/r^3}$ is the Keplerian orbital frequency at radius $r$ and $H$ is the pressure scale height of the gas.

The viscous stress tensor $\bm{\upsigma}$ (following \citet{tassoul-1978}) appears in both the momentum equation \eqref{eq:navier-stokes-momentum} as well as the dissipation function:
\begin{equation}
	\label{eq:dissipation-function}
	Q_\mathrm{visc} = \frac{1}{2\nu\Sigma}\mathrm{Tr}(\bm{\upsigma}^2) = \frac{1}{2\nu\Sigma}\left(\sigma_{rr}^2 + 2\sigma_{r\phi}^2 + \sigma_{\phi\phi}^2 + \sigma_{zz}^2\right),
\end{equation}
where $\nu = \alpha c_\mathrm{s} H$ is the kinematic viscosity parametrized according to the $\alpha$-viscosity model of \cite{shakura_reprint_1973}.
Here, $\alpha$ is a parameter that captures both radial angular momentum transport that leads to accretion onto the star and heating of the disk due to viscous friction. Numerical simulations of (magneto)hydrodynamical instabilities such as the vertical shear instability \citep[VSI,][]{nelson-etal-2013} or the magneto-rotational instability \citep[MRI,][]{balbus-hawley-1991} have provided numerical estimates of $\alpha$, while observations of young stellar objects surrounded by disks have constrained these estimates \citep{dullemond-etal-2018}.
To probe a wide range of diffusion regimes from practically inviscid to moderately viscous, we choose $\alpha\in\{10^{-6}, 10^{-5}, 10^{-4}, 10^{-3}\}$ for our models.

Viscous dissipation leads to the heating of the disk. 
An embedded planet can also deposit significant amounts of thermal energy via the dissipation of spiral shocks \citep{rafikov-2016, ziampras-etal-2020a}. 
As a cooling solution, we allow the disk to relax to a prescribed temperature profile $T_0$ (see, Eq.~\ref{eq:initial-profiles}) over a relaxation timescale $\tau_\mathrm{relax} = \beta/\Omega_\mathrm{K}$ \citep{gammie_nonlinear_2001}.

The thermal relaxation term appears as an additional source term to the energy equation
\begin{equation}
\label{eq:thermal-relaxation}
Q_\mathrm{relax} = -\Sigma c_\mathrm{v}\frac{T-T_0}{\beta}\Omega_\mathrm{K} \Rightarrow \frac{\partial T}{\partial t} = -\frac{T-T_0}{\tau_\mathrm{relax}}
\end{equation}
where $c_\mathrm{v}=\frac{\mathrm{R}_\mathrm{g}}{\mu(\gamma-1)}$ is the heat capacity of the gas at constant volume.
The parameter $\beta$ controls the relaxation timescale, as well as the overall planet--disk interaction process \citep{miranda-rafikov-2020b}, we choose the values $\beta\in\{0.01, 1, 100\}$ which correspond to very fast, moderate, and very slow relaxation.

The gravity of the star and planet are included as a source term in $\vec{g}$. We work in a star-centered coordinate system and embed a planet with mass $M_\mathrm{p}$ at a position $\vec{r}_\mathrm{p}$. 
Thus, the source term reads
\begin{equation}
	\begin{split}
		\label{eq:gravity}
		\vec{g} &= \vec{g}_\star + \vec{g}_\mathrm{p} + \vec{g}_\mathrm{ind} \\
		&=
		-\frac{\mathrm{G}M_\star}{r^3}\vec{r}
		-\frac{\mathrm{G}M_\mathrm{p}}{(d^2 + \epsilon^2)^{3/2}}\vec{d}
		-\frac{\mathrm{G}M_\mathrm{p}}{r_\mathrm{p}^3}\vec{r}_\mathrm{p},
		\quad \vec{d} = \vec{r} - \vec{r}_\mathrm{p}.
	\end{split}
\end{equation}
The terms $\vec{g}_\star$, $\vec{g}_\mathrm{p}$, and $\vec{g}_\mathrm{ind}$ denote the acceleration due to the star, 
the planet, and the indirect term which is a correction needed because the star-centered frame is not an inertial frame.
As we are considering fixed, non-migrating planets, disk feedback on the star and planet is neglected.
The planet's gravitational pull (2nd term in the RHS of Eq.~\eqref{eq:gravity}) is smoothed using a Plummer potential with a smoothing length $\epsilon = 0.6 H(\vec{r})$ that captures the effect of the vertical structure of a more realistic 3D disk \citep{muller_treating_2012} and prevents singularities near the planet's location. 

For simplicity, we do not allow the planet to migrate. 
We chose to limit the degrees of freedom in our model to focus on the dynamics of the vortex and avoid complex and potentially chaotic interplay of the vortex with the planet \citep{lega-etal-2021}.
For the same reason of simplicity, we neglect planetary accretion in our models.

\subsection{Numerics}
\label{sec:numerics}

We use two different codes for our numerical models: \texttt{PLUTO 4.2} \citep{mignone_pluto:_2007}, 
a finite-volume, energy-conserving, shock-capturing code that treats transport by solving the Riemann problem across the interfaces of adjacent cells in both directions ($r$, $\phi$) in an unsplit fashion, 
and our custom \texttt{FARGO} \citep{masset_fargo:_2000} version, \texttt{FargoCPT} \citep{rometsch_migration_2020}, 
which uses a finite-difference, dimensionally-split, second-order upwind method for gas advection. 
Both codes utilize the FARGO method \citep[implemented into \texttt{PLUTO} by][]{mignone_conservative_2012}, in which orbital advection is essentially performed via the Keplerian rotation on top of which the code solves for the residual velocity deviations, significantly relaxing timestep limitations and reducing numerical dissipation in the process \citep{masset_fargo:_2000}. 

The inherent differences between the two numerical schemes make it worthwhile to carry out our simulations using both codes, 
to verify the robustness of our results and test for numerical convergence.
Namely, the strictly energy-conserving nature of \texttt{PLUTO} and the necessity for artificial viscosity to stabilize \texttt{FargoCPT} are discussed in more detail in Sect.~\ref{sec:fargo-vs-pluto-differences}, among others.

\subsubsection{Grid setup}
\label{sec:grid-setup}

Our computational domain spans the full azimuthal extent and a radial range of $r\in[0.2, 5.0]\,r_\mathrm{p} = [1.04, 26.0]\,\mathrm{au}$, with square cells logarithmically spaced so that the cell aspect ratio is preserved. After carrying out a thorough investigation on the effects of our numerical resolution of the recovery of both radial and azimuthal features caused by the planet, we decide to execute our simulations using a resolution of 8 and 16 cells per scale height (hereafter ``cps'') in both directions ($r$, $\phi$).
Because we use a constant aspect ratio together with a logarithmically spaced radial grid, the resolution in cps is constant throughout the domain.

At this resolution, the two codes reach good convergence in terms of the presence and contrast of features shaped by the planet and results agree between the codes.
This translates to a fiducial resolution of $(N_r, N_\phi) = (528, 1024)$ cells for 8 cps, or $(1056, 2048)$ cells for 16 cps.
In addition, using the same resolution in both directions in terms of cps ensures that the effects of numerical viscosity are isotropic (see Appendix~\ref{sec:appendix-grid-resolution}).

\subsubsection{Initial and boundary conditions}
\label{sec:initial-boundary-conditions}

Our disk is initially axisymmetric and in equilibrium in the radial direction, such that the initial radial velocity profile results a constant accretion rate through the disk.
The azimuthal velocity is close to the Keplerian profile, with the correction due to the radial pressure gradient. 
The initial surface density and temperature profiles are simple power laws such that
\begin{equation}
	\label{eq:initial-profiles}
	\Sigma_0(r) = 222\,\mathrm{g}/\mathrm{cm}^2\, \left(\frac{r}{r_\mathrm{p}}\right)^{-1/2},\quad
	T_0(r) = 120.7\,\mathrm{K}\, \left(\frac{r}{r_\mathrm{p}}\right)^{-1},
\end{equation}
with $r_\mathrm{p} = 5.2\,\mathrm{au}$. This temperature profile translates to a disk with a constant aspect ratio $h(r)=H/r = 0.05$. While the general consensus is that protoplanetary disks are flared (i.e., the aspect ratio increases with distance, see for example \citet{dullemond-2000}), we choose to use a constant aspect ratio since the behavior and lifetime of vortices depends on this quantity \citep{hammer_which_2021}. Thus, we can isolate the dependence of vortices on the physical and numerical parameters in our suite of simulations.

The radial and azimuthal velocity components at $t=0$ are then
\begin{equation}
	\label{eq:initial-velocities}
	u_r(r) = -\frac{3}{2}\frac{\nu}{r},\quad
	u_\phi(r) = r\Omega_\mathrm{K}\sqrt{1-1.5\,h^2}
\end{equation}
Near the boundaries, within the radial extent $r\in[0.2,0.25]\cup[4.2,5.0]\,r_\mathrm{p}$, the surface density and velocity are both damped to their initial profiles (see Eqs.~\eqref{eq:initial-profiles}, \eqref{eq:initial-velocities}) using the method of \cite{de_val-borro_comparative_2006}, over a damping timescale of 0.3 periods at the respective boundary.
 While the radial boundary edges are closed, this minimizes the reflection of spiral waves back into the computational domain. The boundaries are periodic in the azimuthal direction.

We then embed a Jupiter-sized planet ($M_\mathrm{p} = 1 \text{M}_\mathrm{J} = 10^{-3} M_\star$) in most models, with some simulations instead containing a less massive planet of $M_\mathrm{p} = 0.5\,\text{M}_\mathrm{J}$. To smoothly introduce the planet into the disk,
we typically allow the planet to grow over 100 orbits at $r_\mathrm{p}$ using the formula by \cite{de_val-borro_comparative_2006}.
The importance of the growth timescale and planet mass will be discussed in Sect.~\ref{sec:results}.

\subsection{Vortex Detection}
\label{sec:vortex_detection}
We use the gas vortensity 
\begin{align}
\label{eqn:vortensity_definition}
\varpi = \frac{(\nabla\times\vec{u}) \cdot\hat{z}}{\Sigma},
\end{align}
where $\hat{z}$ is the unit vector in the vertical direction, as a proxy to detect and track the evolution of vortices over hundreds of snapshots for every model.
Since these vortices consist of anticyclonic motion, the center of a vortex corresponds to a local minimum in vorticity, $\omega = (\nabla\times\vec{u}) \cdot\hat{z}$.
Because vortices tend to accumulate mass towards their center and $\Sigma$ is enhanced inside the vortex, the transition from the background flow to the vortex region is stronger and the vortex is more easily identified in a map of $\varpi$ than in the case of $\omega$ alone.

More precisely, we use the gas vortensity normalized by the background vortensity from the initial conditions, $\varpi_0 = (\nabla\times\vec{u}_\text{K}) \cdot\hat{z} \,/\,\Sigma_0$.
This eliminates the radial dependence of the Keplerian velocity and the disk's surface density
and ensures that our vortex proxy quantity, $\varpi/\varpi_0$, is of order unity everywhere in the disk except for the gap region, due to its very low surface density. The quantity $\varpi/\varpi_0$ usually varies between -1 for strongly counter-rotating vortices to 1 for the background flow.

We use our new Python module, called \texttt{Vortector}, that extracts iso-vortensity contours using the computer vision library \texttt{OpenCV} \citep{opencv_library} to detect vortex candidates and then fits a 2D Gaussian to the vortensity and surface density data. The FWHM (or $2.355\sigma$) of this Gaussian is used to define the radial and azimuthal extent of a vortex.
Using this method, we also extract information about the shape of the vortex, including its radial and azimuthal extent and the mass it encloses.
A more detailed description can be found in Appendix \ref{sec:vortector}.

This automated process has the drawback of sometimes producing detection artifacts, as can be seen for example on Fig.~\ref{fig:ts_rampup} below (top panel, dashed orange line), such that the vortex size (and thus its mass) is overestimated near the end of its lifetime as it blends into the disk background. While this effect is partly counteracted by using a median filter in time, we do not manually edit the output of the \texttt{Vortector} on a model-by-model basis.

\vspace{1cm}
In the following three sections, we present the results of our simulations.
First, we present a typical example of vortex formation and evolution (Sect.~\ref{sec:lifetrack}). 
We then go on to describe the dependence of vortices on physical parameters for the group of vortices with short and intermediate lifetime (Sect.~\ref{sec:results}).
Finally, long-lived and migrating vortices are presented (Sect.~\ref{sec:long-lived}).

\section{Typical life track of a vortex}
\label{sec:lifetrack}

\begin{figure*}[t]
	\centering
	\includegraphics[width=\textwidth]{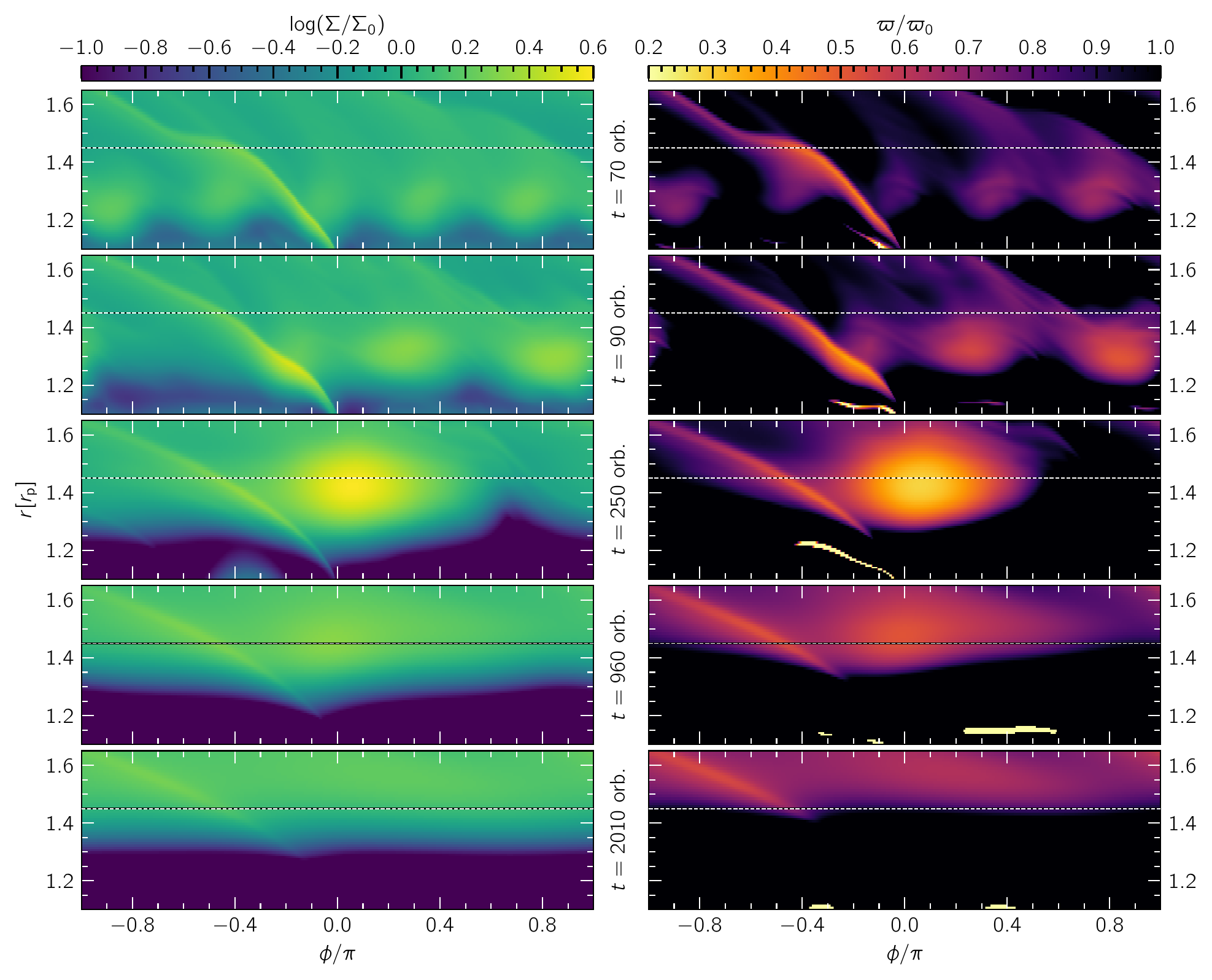}
	\caption{Multiple snapshots of the $\alpha=10^{-5}$, $\beta=1$, 8~cps model showcasing the vortex merging process during the early stage of gap opening, the resulting vortex's fully grown size and its subsequent decay. 
	The surface density and vortensity contrast compared to their initial profiles is shown on the left and right panels, respectively. Time is quoted in units of planetary orbits. 
	The horizontal line at $r=1.45\,r_\mathrm{p}$ serves to highlight the outward radial movement of those structures as the gap around the planet grows wider.
	The planet is located at $r=1\,r_\mathrm{p}$ and $\phi = 0$.}
	\label{fig:example-vortex-heatmap}
\end{figure*}

\begin{figure}[t]
	\centering
	\includegraphics[width=.5\textwidth]{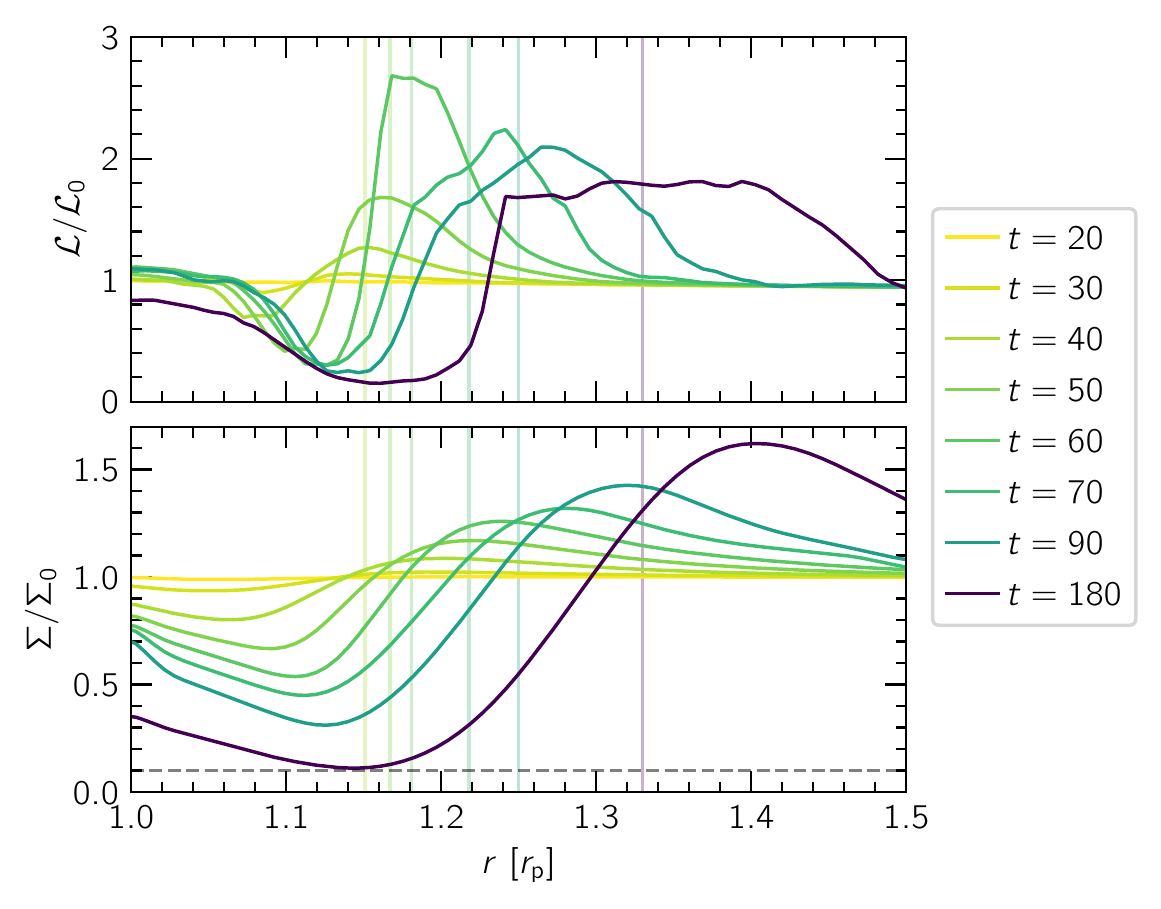}
	\caption{Evolution of radial Lovelace parameter (see Eq.~\eqref{eqn:lovelace_parameter}) and $\Sigma$ profiles during vortex formation over the first 200 orbits of the sample case from Sect.~\ref{sec:lifetrack}.
	The vertical lines indicate the center of the plateau in $\mathcal{L}$ (estimated by eye) to guide the eye to the corresponding location of the $\Sigma$ profile.
	$\mathcal{L}$ is calculated as the azimuthal average at each radius.
	The dotted horizontal line in the bottom panel marks 10\% of $\Sigma_0$ which we define as the location of the gap edge.}
	\label{fig:lovelace-profiles}
\end{figure}
\begin{figure}[t]
	\centering
	\includegraphics[width=.5\textwidth]{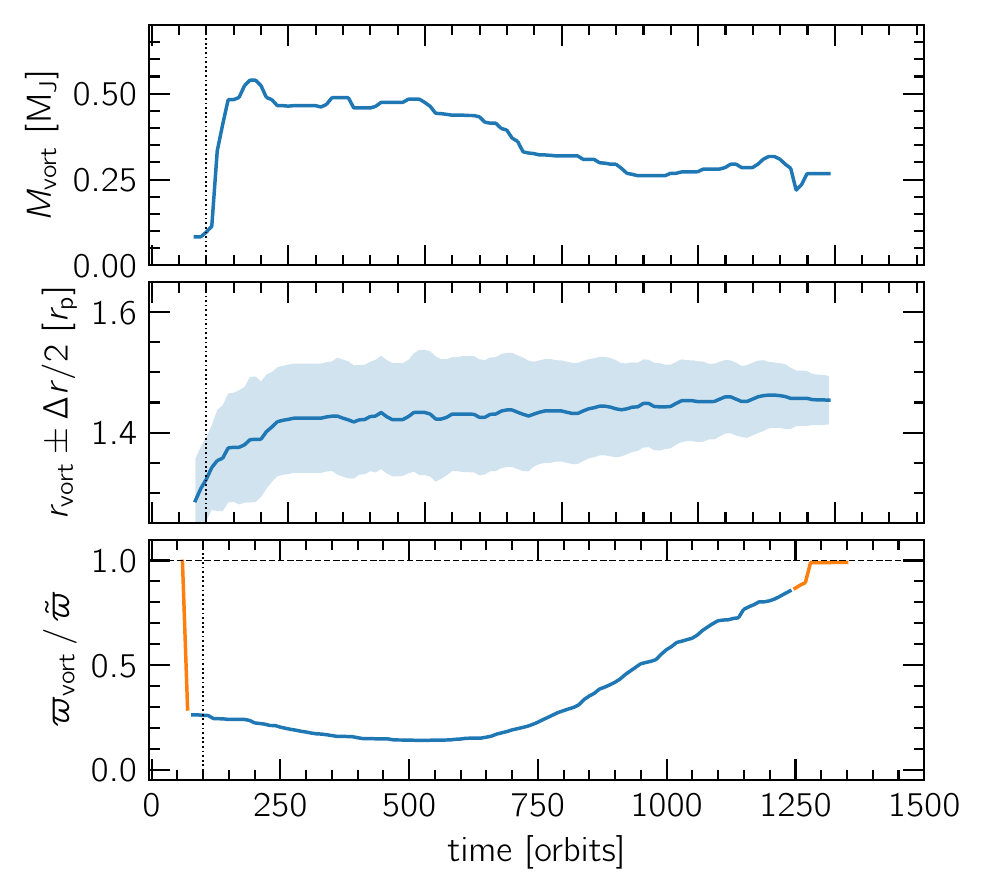}
	\caption{Evolution of vortex properties for the showcase simulations presented in Sect.~\ref{sec:lifetrack}.
	The panels show, from top to bottom, the mass enclosed in the FWHM ellipse of the vortex fit, $M_\text{vort}$ in Jupiter masses, the radial location of the vortex, $r_\text{vort}$ and its FWHM, $\Delta r$, indicated by the shaded area, and the ratio between minimum vortensity inside the vortex and the azimuthal median of vortensity at the radial location of the vortensity minimum.
	A dotted vertical line indicates the time when the planet has reached its final mass.
	The curves are smoothed with a median filter which spans over the next and last 5 datapoints ($\pm50$~orbits at $r_\mathrm{p}$).
	The orange parts of the line in the bottom panel show the evolution of the vortensity prior to the ``birth'' and after the ``death'' of the vortex.
	}
	\label{fig:ts-sample}
\end{figure}

The Jupiter-sized embedded planet opens a deep gap in all of our simulations. 
Fig.~\ref{fig:example-vortex-heatmap} shows maps of $\Sigma$ (left) and $\varpi$ (right) normalized by their initial values at five timestamps during the vortex lifetime for a model with $\alpha=10^{-5}$, $\beta=1$, and a resolution of 8~cps performed with the \texttt{FargoCPT} code.
Horizontal dotted lines at $r= 1.45\,r_\mathrm{p}$ are superimposed as a reference marking the final location of the vortex center.
Here, four small-scale vortices (top row) first merge into two slightly larger vortices (second row) and then finally into one massive vortex (middle) that will last for a little over 1100 orbits.
The vortex slowly decays over time, maintaining a large size (fourth row).
In the later stages, the vortex is no longer present anymore (bottom row). The non-axisymmetric structure still visible exists due to the planet's spiral arm and is corotating with the planet.

During the early gap opening process, the outer gap edge grows Rossby-wave unstable \citep{lovelace_rossby_1999} and several small-scale vortices form around it (top two rows).
Fig.~\ref{fig:lovelace-profiles} shows radial profiles of the Lovelace parameter, $\mathcal{L}$, (top) and $\Sigma$ (bottom) at different timestamps during the vortex formation up until $t = 180$ orbits.
The Lovelace parameter is defined as
\begin{align}\label{eqn:lovelace_parameter}
\mathcal{L} = \frac{S^{2/\gamma}}{\varpi}
\end{align}
with the entropy $S = P/\Sigma^{\gamma}$.
The development of a maximum in $\mathcal{L}$, which is one condition for the onset of the RWI, is visible.
Vertical lines at the center of the $\mathcal{L}$ maxima (as determined by eye) are added to both panels to guide the eye for a comparison of the location of the maxima in $\mathcal{L}$ and $\Sigma$ at each timestamp.
The $\mathcal{L}$ maxima are located on the slope of the gap edge slightly inward of the $\Sigma$ maxima and coincide with the location of the small vortex centers.
The maximum in $\mathcal{L}$ moves outward following the maximum in $\Sigma$ as the gap opens.
This illustrates that the vortices form due to the RWI at the slope of the outer gap edge.

In the absence of self-gravity, these small vortices then quickly merge together (within $\sim$$100$\, planet orbits) into a single large vortex that slowly moves outwards following the gap edge as the gap deepens and widens (third and fourth row in Fig.~\ref{fig:example-vortex-heatmap}). The surviving vortex then typically decays over $\sim$200--2000~orbits.
The evolution of three vortex properties is illustrated in Fig.~\ref{fig:ts-sample}. It shows, from top to bottom, the mass $M_\mathrm{vort}$ enclosed within the FWHM ellipse of the 2D Gaussian fit to $\Sigma$, the location of the center of the vortex $r_\text{vort}$ and the radial FWHM width $\Delta r$ as the shaded area, and the vortensity at the vortex center normalized by the azimuthal median.
The vertical dotted lines indicate the time when the planet reached its full mass (typically 100 orbits).
A short phase of vortex formation is followed by a slow and steady decay process, as can be seen in the decrease of mass and radial size.
Because the vortensity contribution of the anticyclonic vortex is negative, an increase in vortensity indicates a decay as well.
The line in the bottom panel of Fig.~\ref{fig:ts-sample} is continued (in orange) for another 100 orbits after the vortex decayed according to our criterion presented below in Sect.~\ref{sec:results-lifetime}, to illustrate the return of the curve to 1, which corresponds to an azimuthally symmetric state.

During its lifetime, the vortex can become as large as $\Delta r = 0.4\,r_\mathrm{p}$ (2\,au for $r_\mathrm{p} = 5.2\,\mathrm{au}$) with a typical vortex aspect ratio ($r\Delta\phi/\Delta r$) of 6-10.
Its mass, $M_\text{vort}$, is typically some tenths of $M_\text{J}$ but can be as large as one $M_\text{J}$, with a surface density enhanced by a factor of up to 7 compared to the initial value.

The vortices form around the location where the radial $\Sigma$ profile reaches 10\% of its initial value (see bottom panel of Fig.~\ref{fig:lovelace-profiles}), which we define as the gap edge similar to \citet{crida-etal-2006}.
During their lifetime, most vortices tend to stick to this gap edge in the sense that their inner boundary, $r_\text{vort} - \Delta r /2$, roughly coincides with the gap edge location.
For some models, we observe that the vortex detaches from the outer gap edge after several hundred orbits and starts migrating outward.
These models will be discussed later in Sect.\,\ref{sec:long-lived}.

\section{Dependence of vortex properties on physical parameters}
\label{sec:results}

Having described a typical lifetrack of a vortex in our simulations, we now present the effects of different physics and numerics on vortex lifetime, location, and impact on the overall disk structure. 
The model parameters are listed with the main results in Table~\ref{tab:lifetimes}.

\subsection{Vortex lifetime}
\label{sec:results-lifetime}

\begin{figure*}[t]
	\centering
	\includegraphics[width=\linewidth]{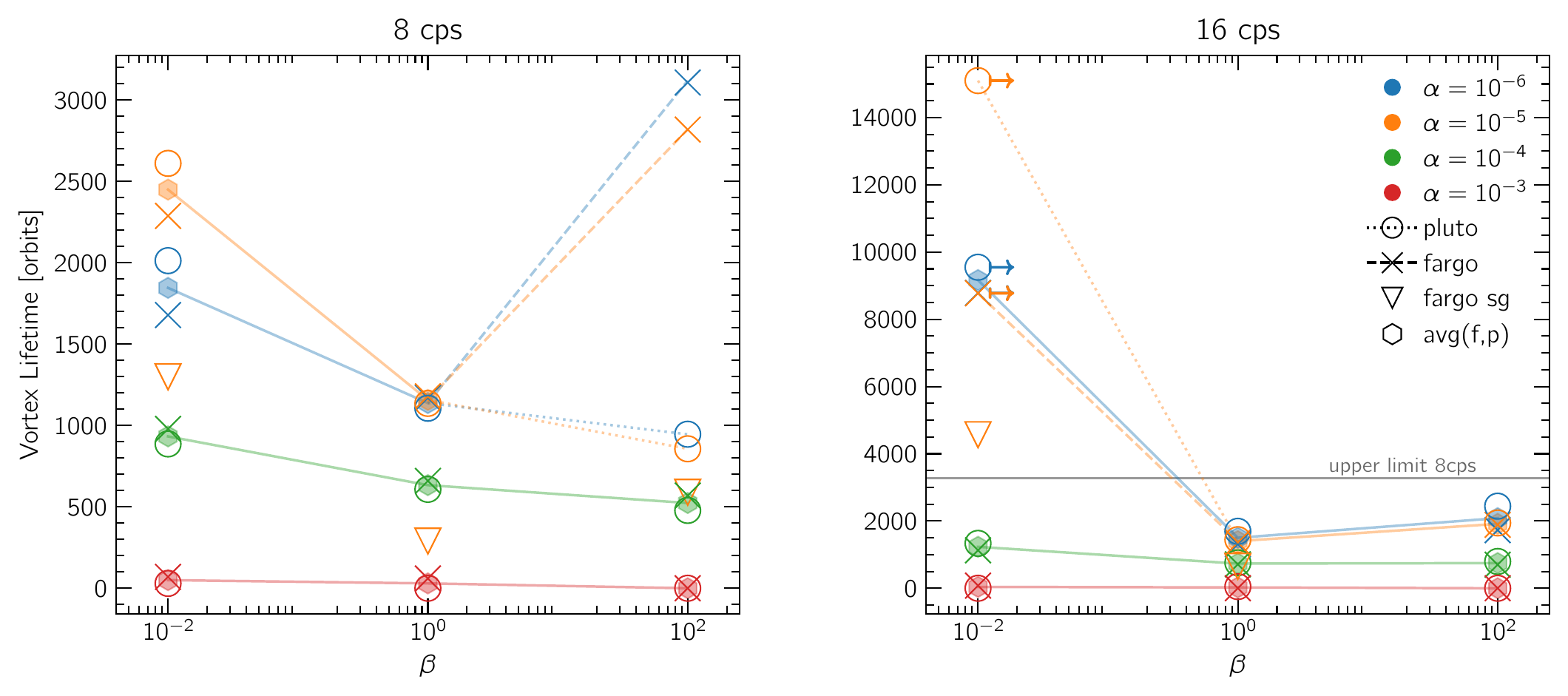}
	\caption{Lifetime of vortices as a function of $\beta$ for 8 cps (left) and 16 cps resolution (right).
	Colors encode $\alpha$, and the different symbols denote the code and the inclusion of self-gravity.
	The solid lines help guide the eye and connect the lifetime averages between the two codes (without self-gravity) for each value of $\alpha$, where the two codes agree sufficiently.
	For parameters where there is a difference between the codes, dashed and dotted lines connect to the datapoints of the \texttt{FargoCPT} and \texttt{PLUTO} runs, respectively.
	A ``$\mapsto$'' next to a symbol marks models that were terminated due to runtime constraints but still contain an active vortex.
	The horizontal gray line in the right panel indicates the top of the y-axis of the left panel.
	A list of all vortex lifetimes shown here is provided in Table~\ref{tab:lifetimes}.
	}
	\label{fig:masterplot}
\end{figure*}

We define the vortex lifetime as the time difference between its ``birth'' and ``death''
by analyzing the ratio of $\varpi$ to the azimuthal median value, $\tilde{\varpi}$ as a function of time.
The normalization with $\tilde{\varpi}$ instead of $\varpi_0$ is done to eliminate the $\varpi$ evolution of the background disk due to changes in $\Sigma$ and radial pressure gradients, which affect $(\nabla \times \vec{u}) \cdot \hat{z}$ by changing the azimuthal velocity.

The ``birth'' is identified as the time when $\varpi/\tilde\varpi$ drops from its initial value of 1 (for an axisymmetric disk) down to lower values (see bottom panel of Fig.~\ref{fig:ts-sample}).
Because $\varpi/\tilde\varpi$ drops already for small vortices, the lifetime also includes the stage where there are multiple small vortices (see Sect.~\ref{sec:lifetrack}).

The ``death'' of the vortex, however, is less obvious to identify.
At this stage, $\varpi/\tilde\varpi$ usually slowly rises back to the background disk's value.
Usually, there is a ``knee'' visible in $\varpi/\tilde\varpi$ at or slightly after the point in time where the vortex dies and where $\varpi$ approaches the background flow (see the orange part of the line in the bottom panel of Fig.~\ref{fig:ts-sample} where $\varpi/\tilde\varpi$ is continued for another 100~orbits after the vortex disappeared at $t = 1250\,\mathrm{orbits}$).
For some models, this ``knee'' is not visible, and we manually inspect the 2D contour plots of $\varpi/\tilde\varpi$ and identify when no closed iso-value lines (with spacing in $\varpi/\tilde\varpi$ of 0.05) are present anymore.
As an additional measure for less obvious cases, we analyze the gas streamlines at different timestamps.

In our models, the drop in $\varpi/\tilde\varpi$ happens in a matter of tens of orbits.
A conservative estimate for the uncertainty of this ``birth'' time measurement is 50 planetary orbits.
From applying the manual method to models where the ``knee'' exists in the $\varpi/\tilde\varpi$ curve (implying the ``death'' of the vortex), we estimate a conservative uncertainty to be 100 planetary orbits.
This leaves a total uncertainty of 150 planetary orbits for the lifetime of our vortices.

The lifetime of vortices in our grid of simulations is shown as an overview in Fig.~\ref{fig:masterplot}.
The left and right panels show vortex lifetimes as a function of $\beta$ for 8 and 16~cps, respectively.
The viscous $\alpha$ is encoded in color, and the symbol indicates the simulation code and the inclusion of self-gravity.
For each value of $\alpha$ and $\beta$, we calculated the average (``avg(f,p)'') between the two codes (not including the self-gravity models) when the results are close together.
The solid-colored lines connect the averages to help visualize the trends.
For parameters for which the two codes showed different vortex lifetimes, we added separate lines connecting the average to the \texttt{FargoCPT} and \texttt{PLUTO} results to highlight the differences.

Lifetimes range from some hundred to 2000 orbits for the shorter-lived vortex group up to at least 15\,000~orbits for the long-lived vortices discussed later in Sect.~\ref{sec:long-lived}.
The most prominent features of the distribution are the trend of decreasing lifetime with increasing $\alpha$ and the minimum of vortex lifetime at $\beta=1$ for low $\alpha$ and high resolution.

In the following sections, we address the influence of our model parameters on vortex lifetime.

\subsection{Influence of the thermal relaxation timescale}
\label{sec:lifetime-beta}

The dimensionless thermal relaxation timescale $\beta$ has a strong effect on vortex lifetime.
For $\alpha = 10^{-4}$, lifetimes are of the order of several hundred to 1000~orbits with a downward trend as $\beta$ increases.
Vortex lifetimes are shortest for $\beta=1$ (around 1250~orbits) and increase towards both sides to around 2000~orbits for $\beta = 100$ and to values of the order of 10\,000~orbits for $\beta=0.01$, high-resolution runs. This decrease in lifetime for non-isothermal disks is consistent with the results of \citet{tarczay-nehez_vortex_2020}. Exceptions to this trend are the 8-cps \texttt{PLUTO} models for $\alpha=10^{-5}$--$10^{-6}$ and $\beta=100$. We could not identify the reason why the two codes did not agree for these parameters, but we note that the two codes match well once again for 16~cps in the same configurations.
Models with very long vortex lifetimes will be analyzed later in Sect.~\ref{sec:long-lived}.

\citet{fung_cooling-induced_2021} reported a similar trend in vortex lifetime in two-dimensional shearing-box simulations without planets, in which the vortex was introduced by initializing the simulation with a radial density bump.
They found that vortex decay is fastest for intermediate $\beta$ in the range 1--10, but their disk model assumes a constant background disk, without gradients in $T$ and $\Sigma$ which change baroclinic effects.
Our results indicate that a similar mechanism might be at play in the presence of an embedded planet with strong spiral arm shocks.
However, the strong enhancement of vortex lifetime for $\beta = 0.01$ hints at the presence of an additional mechanism which keeps the vortices alive. 
We discuss these hypotheses further in Sect.~\ref{sec:vortex-conditions}.

\begin{figure}[t]
	\centering
	\includegraphics[width=0.5\textwidth]{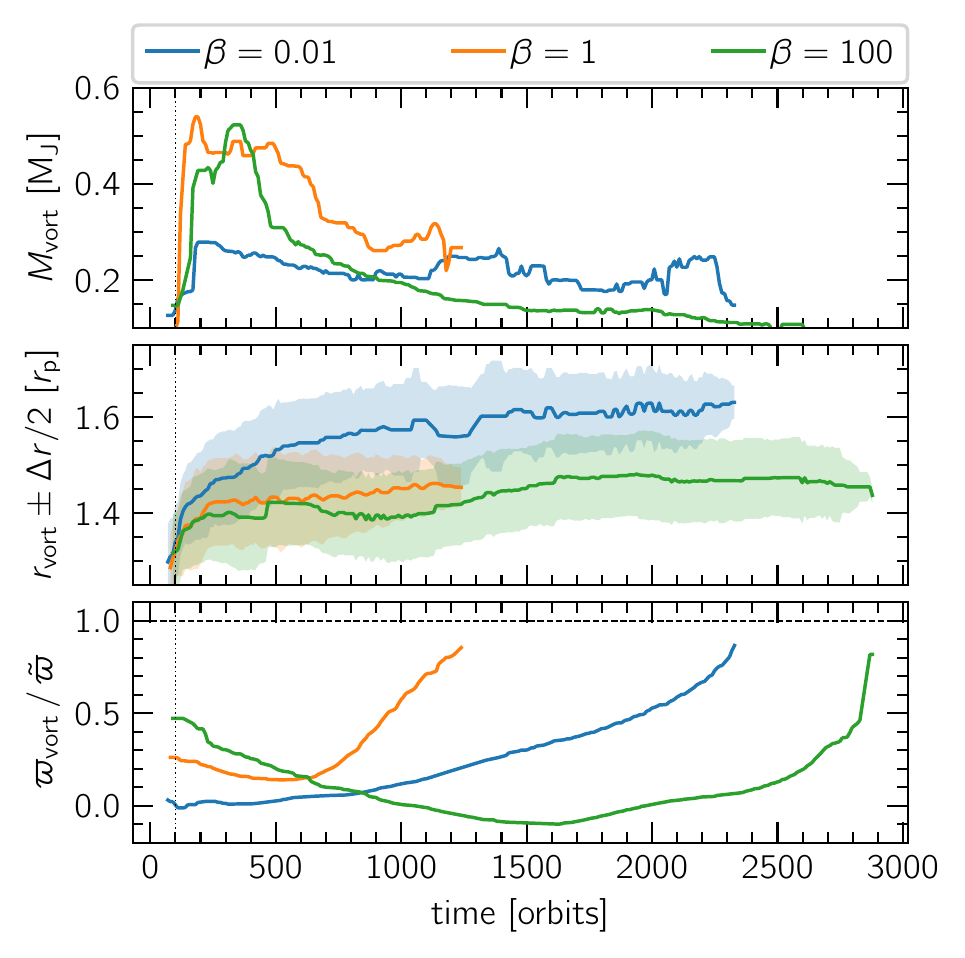}
	\caption{Evolution of vortex properties for varying values of the thermal relaxation $\beta$ parameter.
	The panels are as in Fig.~\ref{fig:ts-sample}.
	Shown are models run with the \texttt{FargoCPT} code with $\alpha = 10^{-5}$ and at 8~cps resolution (orange ``f8'' dots in Fig.~\ref{fig:masterplot}).
	\label{fig:ts_beta}
	}
\end{figure}

For a comparison of vortex evolution at different $\beta$ see Fig.~\ref{fig:ts_beta}, where the evolution of vortex properties (analogous to Fig.~\ref{fig:ts-sample}) of three \texttt{FargoCPT} simulations at 8~cps resolution with $\beta = 0.01, 1, 100$ is shown.
The absolute radial location of vortices varies with $\beta$, as well.
This is due to the tendency of the vortices to form and subsequently stick to the outer planet gap edge and the gap opening process being strongly influenced by $\beta$.
\citet{miranda-rafikov-2020b} showed that ``extreme'' values of $\beta$ (i.e., $\beta\rightarrow0$ or $\beta\rightarrow\infty$) result in narrower planet-opened gaps but additional gaps in the inner disk, 
whereas intermediate values of $\beta\sim 0.1$--10 lead to a single, wide gap around the planet's orbit. 
In our simulations, models with $\beta=0.01$ show the widest gaps, narrower gaps are present for $\beta=1$, and $\beta=100$ models showed an even slightly narrower gap.
This is reflected in the vortex locations which are further in for higher $\beta$ (see the center panel of Fig.~\ref{fig:ts_beta}).
The difference to \citet{miranda-rafikov-2020b} might be due to the presence of the vortex.

\subsection{Planet growth timescale}

\begin{figure}[t]
	\centering
	\includegraphics[width=0.5\textwidth]{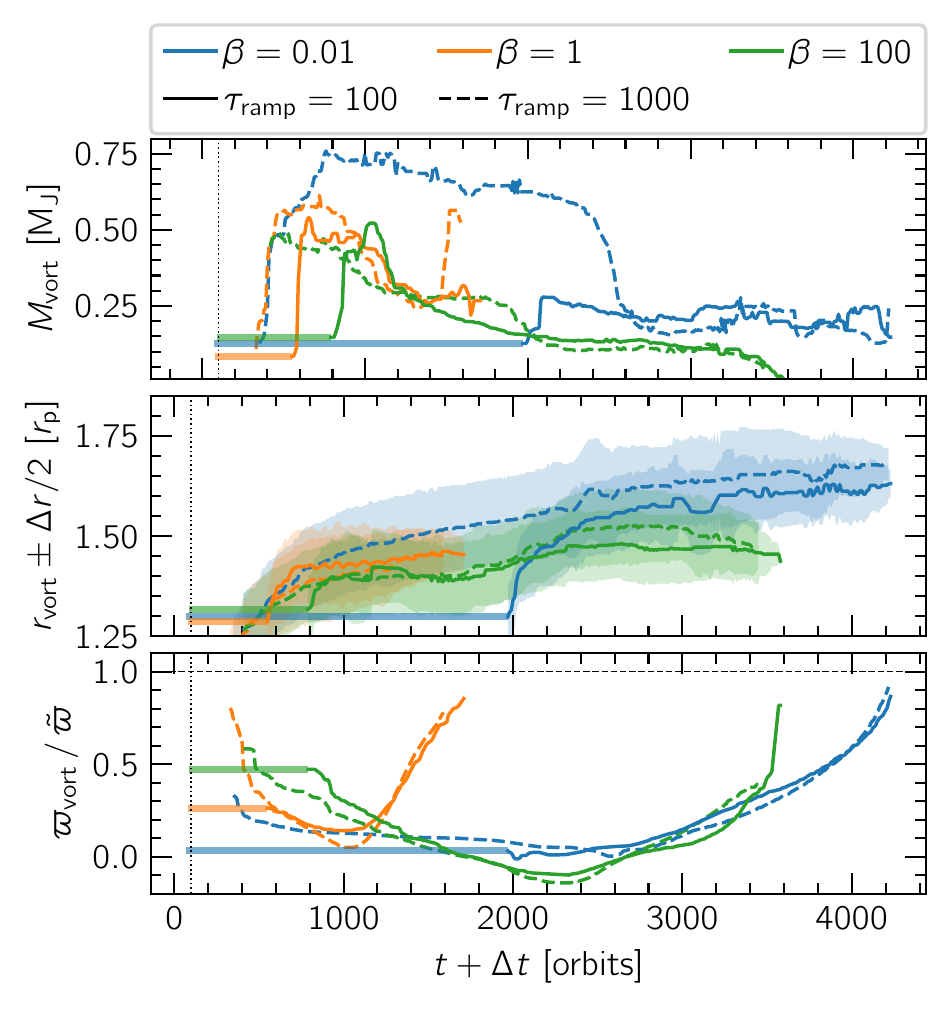}
	\caption{Influence of the planet introduction time on the evolution of vortex properties.
	The panels are as in Fig.~\ref{fig:ts-sample}.
	Solid and dashed lines show models with a $\tau_\mathrm{ramp} = 100\,\mathrm{orbits}$ and $1000\,\mathrm{orbits}$, respectively.
	The $\tau_\mathrm{ramp} = 100\,\mathrm{orbits}$ curves are shifted to the right (see the horizontal lines) to illustrate that the curves have the same shape in the decay phase, independent of $\tau_\mathrm{ramp}$.
	Note that the final evolution of the vortex, after it has reached its minimum in vortensity, is the same independent of planet introduction time.
	\label{fig:ts_rampup}
	}
\end{figure}

\cite{hammer_slowly-growing_2017} observed that the lifetime of planet-induced vortices can depend on the timescale over which the planet mass is
increased in order to introduce the planet into the simulation.
They found that vortex lifetime decreased with a longer planet growth time.
In our models, increasing the planet growth timescale from 100 to 1000 orbits caused vortices to live longer by 470 for $\beta=1$ up to 1900 orbits for $\beta=0.01$.
Fig.~\ref{fig:ts_rampup} shows the evolution of vortex quantities comparing the \texttt{FargoCPT} runs 
with a $\tau_\mathrm{ramp} = 100\,\mathrm{orbits}$, already presented in Fig.~\ref{fig:ts_beta}, 
with their respective counterparts with $\tau_\mathrm{ramp} = 1000\,\mathrm{orbits}$.
The curves of runs with $\tau_\mathrm{ramp} = 100\,\mathrm{orbits}$ are shifted to the right by a time $\Delta t$ from 470 to 1900 orbits.
This shift clearly illustrates that the decay of these vortices is almost the same for both values of $\tau_\mathrm{ramp}$
in terms of their mass, location and vortensity curves.
The only difference caused by the planet injection timescale is how long it takes for the vortex to reach
the turnover point, after which it starts to decay.

\subsection{Planet mass}
\label{sec:results_planet_mass}

From our $M_\text{p} = 0.5\,\text{M}_\text{J}$ models we can not draw any conclusions regarding the dependence of vortex lifetime
on planet mass, because for the set of parameters, $\beta=0.01$ and $\alpha=10^{-5}$, the vortices 
are long-lived outliers like the ones discussed in Sect.~\ref{sec:long-lived}.
However, the location of the vortex is also influenced by the planet's mass. 
Lower-mass planets open narrower gaps and cause the location of the vortex, given that this is linked to the gap edge's location, to be further in compared to more massive planets.
In our models, the vortices in the $M_\text{p} = 0.5\,\text{M}_\text{J}$ were located $\sim 0.15\,r_\mathrm{p}$ closer to the star.

\subsection{Viscosity}
\label{sec:viscosity}

The observed vortex lifetime typically increases with lower values of $\alpha$.
Simulations with $\alpha = 10^{-3}$ show only small vortices forming. They disappear within 100 orbits, 
thus, they are already gone by the time the planet has grown to its full mass.
For models with $\alpha=10^{-4}$, we observe vortex lifetimes of up to around 1000 orbits.

\begin{figure}[t]
	\centering
	\includegraphics[width=0.5\textwidth]{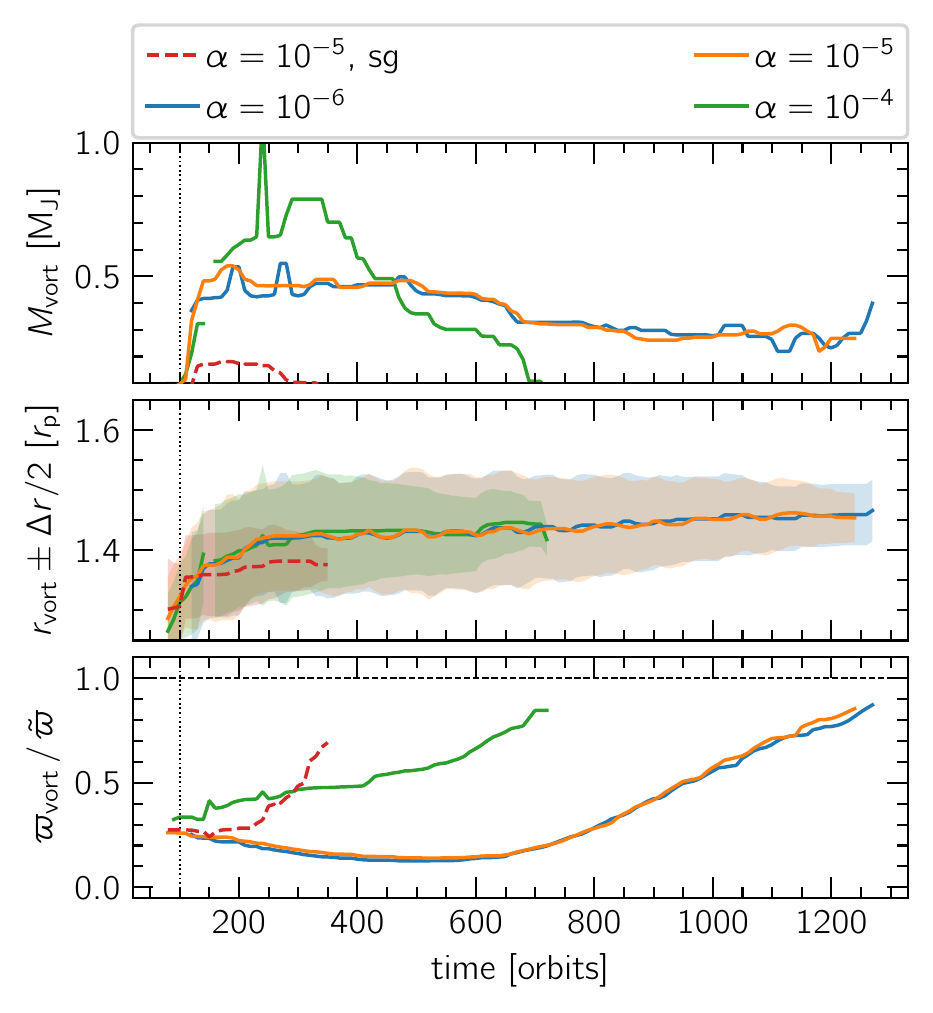}
	\caption{Evolution of vortex properties for varying values of $\alpha$.
	Panels are shown as in Fig.~\ref{fig:ts-sample}.
	Shown are models run with the \texttt{FargoCPT} code with $\beta=1$ and at 8~cps resolution.
	The $\alpha=10^{-3}$ run is excluded because no vortex forms.
	In addition, a run with disk self-gravity enabled is added for the $\alpha=10^{-5}$ case.
	The similarity between simulations with $\alpha=10^{-5}$ and $10^{-6}$ is apparent.
	\label{fig:ts_viscosity}
	}
\end{figure}

Simulations with a lower viscosity ($\alpha=10^{-6}$--$10^{-5}$) show even longer lifetimes, 
usually in the range between 1000 and 2000 orbits, excluding the outliers that we discuss later in Sect.~\ref{sec:long-lived}.
For this range of $\alpha$, vortices usually have similar lifetimes for simulations sharing the same $\beta$ value.
For an example, see Fig.~\ref{fig:ts_viscosity} which shows, from top to bottom, the evolution of the mass enclosed in the vortex's region (FWHM), the location and radial extent (in FWHM) of the vortex as determined by the surface density fit, and the ratio of normalized vortensity to the azimuthal median of the latter at the location of the vortex.

The vortex location is not influenced by viscosity.
Although the gap opening time is 
$t_\mathrm{gap} \approx 2700 \left(\frac{\alpha}{10^{-3}}\right)^{-3/2}\,T_\mathrm{orb}$ according to the estimate in \cite{kanagawa_modelling_2017}, the bulk of the gas in the planet's vicinity is cleared within the first few hundred orbits.
During this time, $\Sigma$ is lowered by two orders of magnitude within the gap region, and the radial gradient of $\Sigma$ becomes steep enough to facilitate vortex formation.

\subsection{Self-gravity}
\label{sec:selfgravity}

Several studies showed that vortices in weakly or strongly self-gravitating disks might not grow as large because small vortices do not merge into one large vortex \citep{lin_effect_2011} and dissipate more rapidly due to stretching in the azimuthal direction \citep{lovelace_rossby_2013,regaly_vortex_2017,zhu_gas_2016}.
This can be the case even for low-mass disks as long as the Toomre stability parameter $Q$ is lower than 50 or $h Q \leqslant \frac{\pi}{2}$.
For the choice of parameters in our models, the Toomre parameter is $Q\approx25\,(r/r_\mathrm{p})^{-3/2}$ ($hQ \approx 1.25\,(r/r_\mathrm{p})^{-3/2}$), 
dropping under 5 at roughly $r/r_\mathrm{p}=2.8$.
To check the effect that disk self-gravity has played in our models, 
we ran additional simulations with \texttt{FargoCPT} with self-gravity activated for all three values of $\beta = 0.01, 1, 100$ and for $\alpha = 10^{-5}$.

The lifetimes of vortices in these simulations are shown in Fig.~\ref{fig:masterplot} as the rightmost datapoint in each column (models ``f8sg''). An example evolution of their properties is shown in Fig.~\ref{fig:ts_viscosity}.

Self-gravity inhibits the merging of the small initially formed vortices in to one large vortex.
Instead, two smaller vortices usually remain until they decay.
This leads to a significantly shorter lifetime compared to the analogous simulations without self-gravity, consistent with the above studies.
This, however, does not apply to the long-lived, migrating vortices which are discussed in Sect.~\ref{sec:long-lived}.

Fig.~\ref{fig:ts_viscosity} shows that the center of the vortex in a model with self-gravity and $\beta=1$ is further in compared to its non-self-gravitating counterpart.
This is due to a smaller radial extent of the vortex in the run with self-gravity and the tendency of the inner edge of each vortex to coincide with the gap edge.

Because self-gravity does not noticeably change the radial disk profile for the mass regime of our models, the inner edge of the vortices is at the same location, independent of whether self-gravity is included or not.
The same effect is also observed for $\beta = 0.01$ and $\beta = 100$.

\section{Long-lived and migrating vortices}
\label{sec:long-lived}

\begin{figure}[t]
	\centering
	\includegraphics[width=0.5\textwidth]{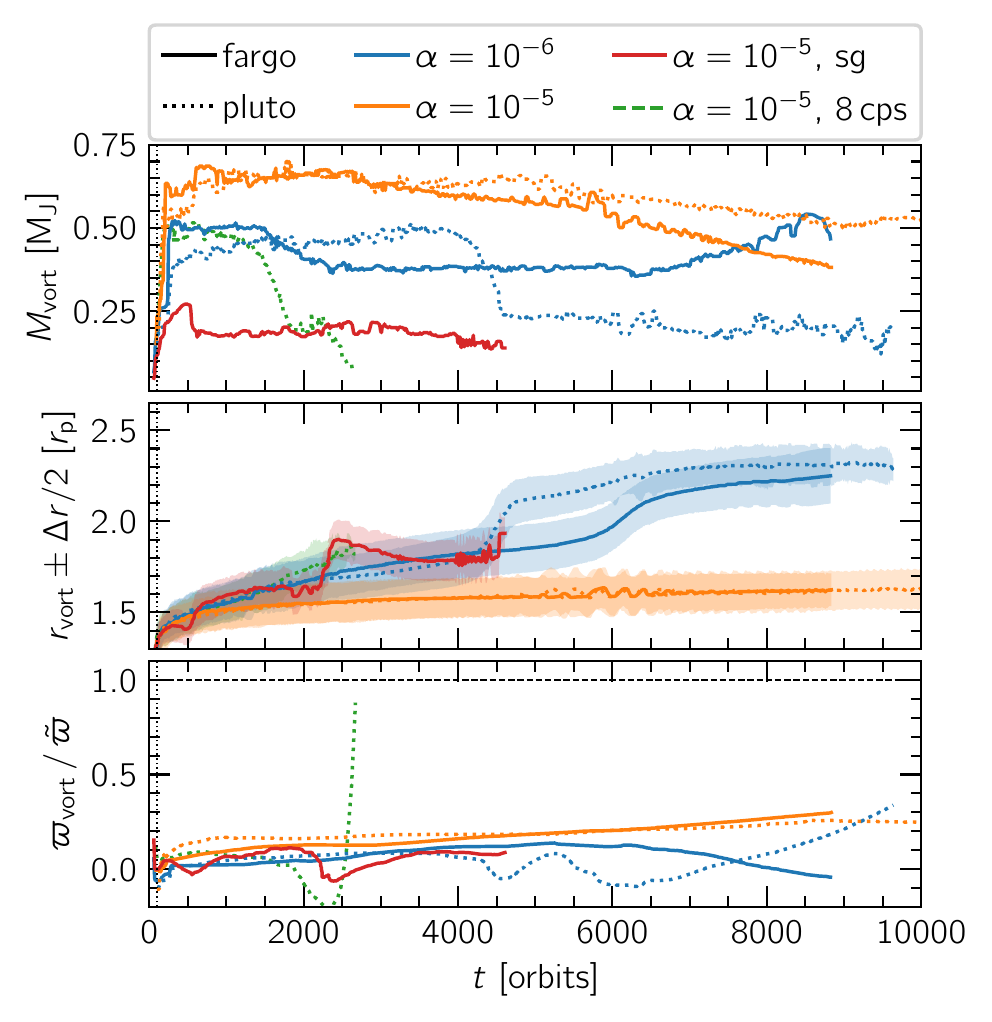}
	\caption{Selection of models with long-living and migrating vortices at 16~cps resolution for the two different codes.
	Both codes agree remarkably well for the blue and orange cases.
	The panels are as in Fig.~\ref{fig:ts-sample}.
	In  models shown here, $\beta=0.01$. The values of $\alpha$, resolution and code used are indicated in the legend.
	\label{fig:ts_outliers}
	}
\end{figure}

\begin{figure*}[th]
	\centering
	\includegraphics[width=\textwidth]{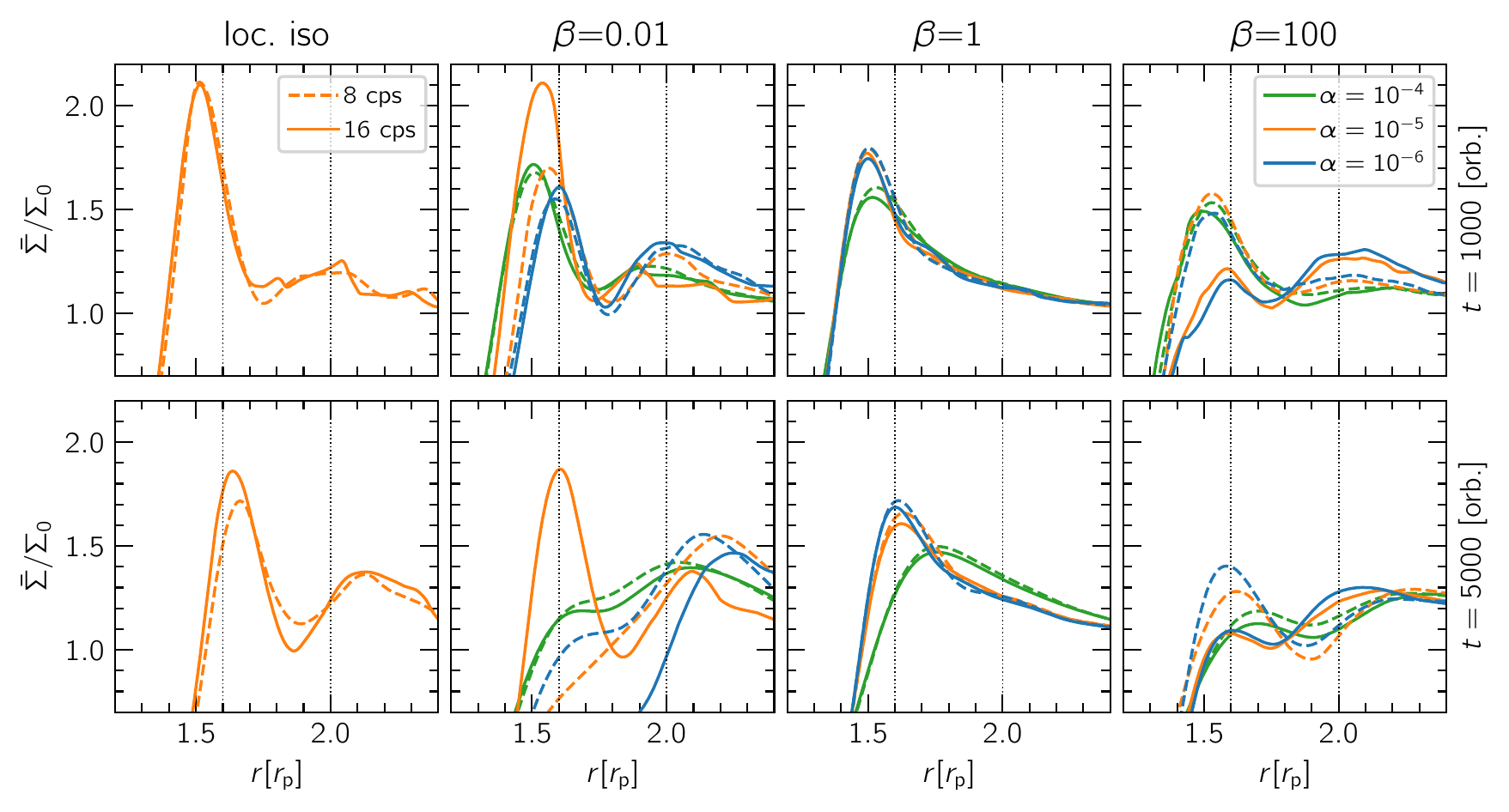}
	\caption{Azimuthally averaged surface density profiles as a function of different physical ($\alpha$, $\beta$) and numerical (cps) parameters at two different timestamps. 
	The peak around $r/r_\text{p}=1.5$--$1.6$ corresponds to the pressure bump formed by the planet as the latter pushes material away, forming a gap around its orbit. 
	The smaller, secondary peak at around $r/r_\text{p}=2.1$ is caused by the vortex that forms near the ``primary'', planet-generated bump. 
	Top: radial profiles at $t=1000$~orbits. At this stage, all models pictured feature a vortex near the primary bump. We note the absence of a secondary bump for the models with $\beta=1$. Bottom: the same profiles at $t=5000$ orbits. 
	Here, the primary bump has moved radially outwards as the planet's gap gets deeper and wider. 
	We highlight the depletion of gas near the ``primary'' pressure bump for the second panel from the left ($\beta=0.01$). 
	This is caused by the combination of a vortex migrating outwards to the secondary bump, and the inability of the planet to resupply that zone with material from its now-depleted gap region. 
	Also note the difference between resolutions of 8 and 16~cps (dashed and solid lines), especially for the $\beta=100$ models and the $\beta = 0.01,\,\alpha=10^{-5}$ model (the evolution of this model is shown as the orange line in Fig.~\ref{fig:ts_outliers}).}
	\label{fig:gap-edges}
\end{figure*}

In some of the cases, a much longer-lived vortex is observed.
In these models, vortices stay close to their peak mass for several thousand orbits and, in some cases, migrate outwards after having stayed at the planet gap edge.
This happens only for very low viscosities ($\alpha\leq10^{-5}$) and $\beta = 0.01$ or locally isothermal simulations ($\beta \to 0$).
For our standard $M_\text{p}=1\,\text{M}_\text{J}$ planets, the long-lived outliers appear only at the highest resolution of 16~cps but not at 8~cps.
For the corresponding $M_\text{p}=0.5\,\text{M}_\text{J}$ model, the long-lived vortex also appeared at 8~cps.
Spiral arms launched by the vortex are clearly visible for these long-lived large vortices (see Fig.~\ref{fig:vortector_overview}).
They are more pronounced for lower values of $\alpha$.

Figure~\ref{fig:ts_outliers} shows the evolution of vortex properties for a selection of models to highlight the observed behavior.
The most prominent example is the model with $\alpha=10^{-5}$, $\beta = 0.01$ and a 16~cps resolution.
The vortex in those runs lived for 15\,100 orbits before we terminated the two simulations due to their long runtime.
Both codes, \texttt{PLUTO} and \texttt{FargoCPT}, agree well for the long-lived cases. 
Specifically, they are in exceptionally close agreement for $\alpha=10^{-5}$ and only differ at later stages for $\alpha=10^{-6}$ (see orange and blue lines in Fig.~\ref{sec:long-lived}).

We do not currently fully understand the mechanism that allows these long-lived vortices to sustain themselves for such long timescales. We attempt to provide a speculative explanation in Sect.~\ref{sec:vortex-conditions}.

For $\beta \not=1$, a secondary radial density and pressure bump is observed in the outer disk.
This is the result of the vortex generating spiral arms which transport angular momentum.
Radially outwards, this results in the accumulation of mass in a second bump 
(see panels for $\beta \not= 1$ in Fig.~\ref{fig:gap-edges}).
This does not happen for $\beta = 1$ due to the less efficient angular momentum transport by spiral arms
for this intermediate value of $\beta$ \citep{miranda-rafikov-2020b}.

For $\beta = 0.01$, some models show vortices migrating radially outwards (e.g., the $\alpha=10^{-6}$ models in Fig.~\ref{fig:ts_outliers}).
This is likely related to the formation of the secondary bump outside of the vortex location (see Fig.~\ref{fig:gap-edges}) and the fact that vortices typically migrate towards pressure bumps \citep{paardekooper_vortex_2010}.

For vortices that migrate far enough outside, which only happens for $\beta=0.01$, 
a weaker secondary vortex appears between them and the planet's gap edge (see Fig.~\ref{fig:vortector_overview}).
These secondary vortices then decay over a few hundreds of orbits, already having decayed by the time the ``primary'' vortex disappears.
While they are treated as independent entities, they are not included in Fig.~\ref{fig:masterplot} or the discussion above. 

Their occurrence is likely the result of a multistage process which begins with the secondary bump forming and the ``primary''
vortex migrating radially outwards towards it and meanwhile supplying mass towards the planet-generated gap edge.
This then feeds the emerging ``secondary'' vortex.

\section{Discussion}
\label{sec:discussion}

In this section, we address some ways in which our results could be interpreted and their relevance in explaining observations.
We also underline some caveats of our models.

\subsection{On the conditions to form and sustain a vortex}
\label{sec:vortex-conditions}

To form a vortex, one needs to create a local vortensity extremum.
In the absence of non-conservative forces, the evolution equation for the vortensity in a two-dimensional flow reads
\begin{align}\label{eqn:vortensity}
	\frac{\partial \varpi}{\partial t} + \vec{u}\cdot \nabla \varpi = \frac{\nabla \Sigma \times \nabla P}{\Sigma^3} \cdot \hat{z} + \mathcal{V} = \mathcal{B} + \mathcal{V}
\end{align}
where $\mathcal{B} = \frac{\nabla \Sigma \times \nabla P}{\Sigma^3} \cdot \hat{z}$ is the baroclinic term and $\mathcal{V}$ describes viscous diffusion of vortensity which can lead to vortex decay.

As outlined in the introduction, several instabilities have been discovered that provide a mechanism to form or destroy large-scale vortices, but they all fundamentally rely on Eq.~\eqref{eqn:vortensity} to change the vortensity of the flow.
The mechanism responsible for the formation of the vortex in our simulations is most likely the RWI which is triggered during the gap opening process, as we demonstrated in Sect.~\ref{sec:lifetrack} and Fig.~\ref{fig:lovelace-profiles}.

To check whether vortices can only form during the gap opening process and not in the quasi-steady state after the bulk of the gas has been pushed out of the gap region, we removed the long-lived vortex from the $\alpha=10^{-5}$, $\beta=0.01$, 16~cps model by replacing the velocities and $\Sigma$ with their azimuthal median values for $r>r_\mathrm{p}$ during the peak of its activity ($t=1880$ orbits).
The fact that there is no vortex forming again is an indication that the formation of vortices in our simulations depends on the gap opening process to produce conditions that can trigger the RWI.
This is also backed by the observation that the peak in $\mathcal{L}$ is strongest for an intermediate time, $t=70$~orbits, during the gap opening process, after which the maximum disappears and a plateau in $\mathcal{L}$ forms.

Vortex decay happens due to at least two mechanisms.
Viscous spreading attacks the vortices for high $\alpha = 10^{-4}$--$10^{-3}$, as illustrated by the trend of lower vortex lifetime for higher $\alpha$, and vortex stretching due to self-gravity effects additionally limits vortex lifetime if it is considered \citep{lin_effect_2011, zhu_gas_2016, regaly_vortex_2017}.
For sufficiently low $\alpha$, another process that depends on $\beta$ starts to be dominant.
We do not fully understand the mechanism but we observed some similarities to the recent work by \citet{fung_cooling-induced_2021}.
They found that, in their simulations, vortices decay the fastest  for $\beta = 1$--$10$ and decay slower for both smaller and larger $\beta$.
Vortex lifetime in their simulations changed by up to an order of magnitude depending on $\beta$.
We also find a minimum in vortex lifetime for $\beta = 1$ with lifetimes increasing as $\beta \not= 1$.

\begin{figure}[t]
	\centering
	\includegraphics[width=.5\textwidth]{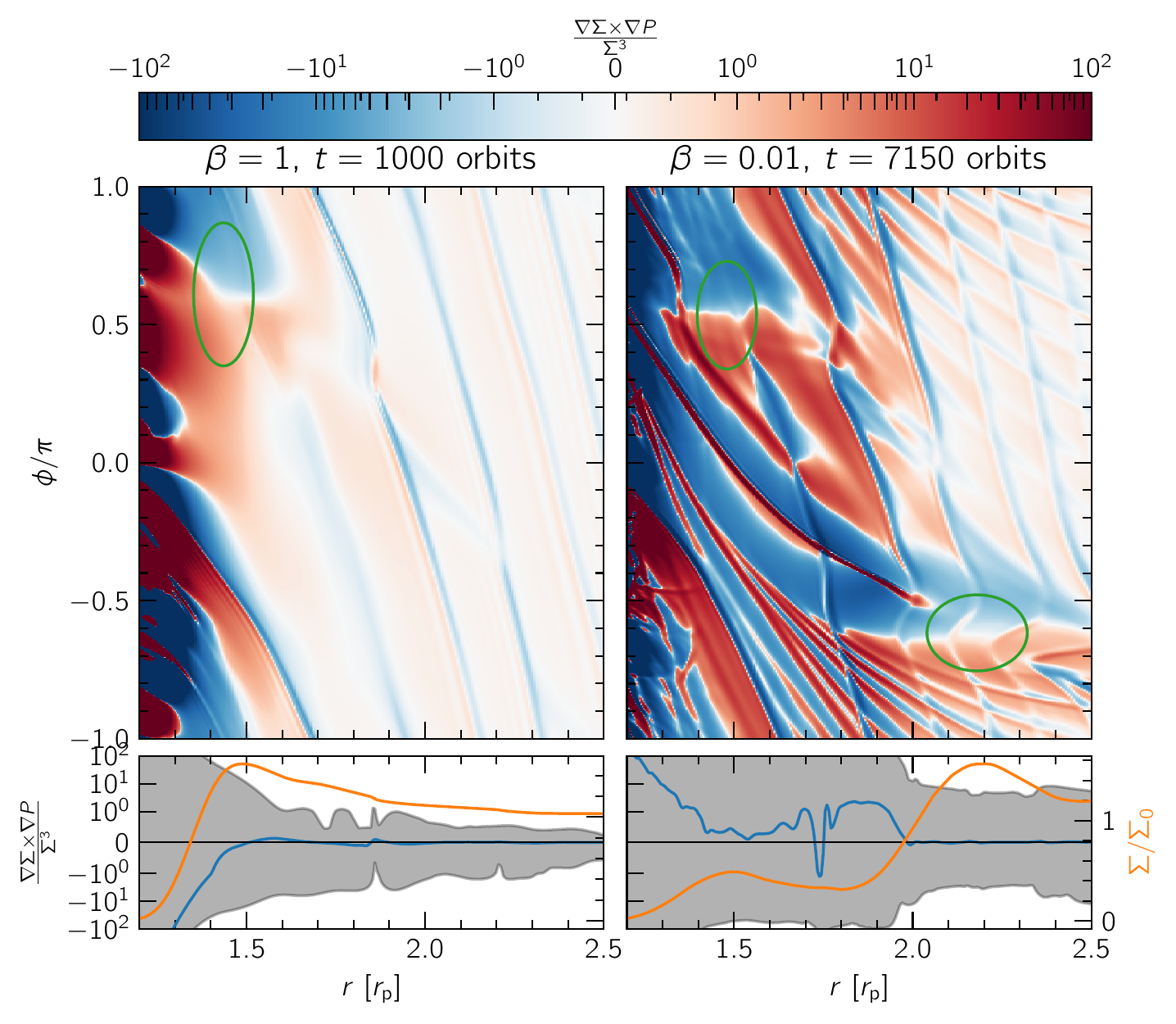}
	\caption{
		Baroclinic term (RHS of Eq.~\eqref{fig:baroclinic}) in the outer disk for a short-lived model ($\beta = 1$ at $t=1000$~orbits) and a long-lived vortex ($\beta = 0.01$ at $t=7150$~orbits, see also Fig.~\ref{fig:vortector_overview}) with $\alpha=10^{-5}$ and 16~cps resolution on the left and right side, respectively.
		The top row shows maps of the baroclinic term with the detected vortices indicated with green ellipses as obtained from the $\Sigma$ fit.
		The bottom row shows the radial $\Sigma$ profile in orange, the azimuthally averaged baroclinic term in blue and the region between its minimum and maximum shaded in gray.
	}
	\label{fig:baroclinic}
\end{figure}

\citet{fung_cooling-induced_2021} explained the decay mechanism by asymmetries in the structure of $\mathcal{B}$ around the vortex center, which they found to be quadrupolar (see their Fig. 6) and to change with $\beta$.
We also find asymmetries in the structure of $\mathcal{B}$, however our simulations differ from theirs in some fundamental aspects.
Our simulations are global with radially varying $\Sigma$ and $T$ profiles and include a planet that continually perturbs the disk, whereas their simulations consider a local shearing sheet with a constant background $\Sigma$ and $T$, with only an initial perturbation in the form of a density bump.
As a consequence of the radially varying $T$ in our simulations, the structure of $\mathcal{B}$ around the vortex center is dipolar in the azimuthal direction, as can be expected for a Gaussian-like density maximum.
Additionally, the planetary spiral arms strongly influence $\mathcal{B}$.
Figure~\ref{fig:baroclinic} shows a 2D map of $\mathcal{B}$ for two simulations with $\alpha=10^{-5}$ and 16~cps resolution.
The left panels show a short-lived vortex with $\beta=1$ at $t=1000$~orbits and the right panels show the long-lived vortex model which exhibits the ``secondary'' vortex (see Sect.~\ref{sec:long-lived} for a description and Fig.~\ref{fig:vortector_overview} for $\varpi$ and $\Sigma$ maps at the same time).
The actual shape of the perturbation of $\mathcal{B}$ inside and around the vortex varies in time because it depends on the phase w.r.t.~to the spiral arm.
It is not clear to us, at the moment, how the changes in structure of $\mathcal{B}$ lead to the change in vortex decay and how this proposed mechanism depends on the various parameters in our system.

The long-lived group of vortices for low $\beta$ (see Sect.~\ref{sec:long-lived}) indicates that there might be another vortex formation mechanism at play. 
Given that the RWI already caused finite perturbations in the disk and our disks exhibit a radial entropy gradient, the SBI \citep{klahr_turbulence_2003,lesur_subcritical_2010} seems to be a natural candidate.
However, we verified that the SBI is not active in our disks by analyzing the Richardson number, the ratio of the buoyancy (also called Brunt-Väisälä) frequency to the shear rate, which needs to be negative in a radially extended region over the full azimuth of the disk for the SBI to operate.
The Richardson number in our simulation is positive, except for narrow stripes following the spiral arms, which rules out that the SBI is active.

To rule out that the difference in lifetime is a result of the initial vortex formation during gap opening, we took the long-lived vortex out of the $\alpha=10^{-5}$, $\beta=0.01$, 16~cps model and inserted it into the 
$\alpha=10^{-5}$, $\beta=1$, 16~cps model.
Although this artificial vortex has the same structure as in its original $\beta=0.01$ model, it decays over nearly the same time as the standard $\beta=1$ vortex.
This is an additional indication that the difference in lifetime is caused by the dependence of the decay process on $\beta$ or a possible additional vortex formation channel that sustains the vortex at low $\beta$.

This leaves us with the hypothesis that the interaction of the spiral arms with the vortices might play a major role in either slowing down vortex decay or providing an additional vortex formation channel.
This hypothesis is motivated by the strong impact of the spiral arms on $\mathcal{B}$ and the dependence of spiral arm properties on $\beta$ \citep{ziampras-etal-2020b,miranda-rafikov-2020b}.
Another contribution might be the vortensity jump across the spiral arm shock, which was recently illustrated to be important for the evolution of vortensity in the case of sub-thermal-mass planets \citep{cimerman_planet-driven_2021}.
Providing an analysis of both mechanisms in our context is, unfortunately, out of the scope of the present explorative study.

\subsection{Effect of in-plane radiation transport}
\label{sec:FLD}

It has been shown that parametrizing radiative effects with $\beta$ while omitting the effects of in-plane radiation transport can result in a potentially inaccurate radial surface density structure mainly in the inner disk and around the gap, due to the impact of $\beta$ on the capability of a planet to open multiple ``secondary'' gaps at $r<r_\mathrm{p}$ \citep{miranda-rafikov-2020b}. 
Here, we are not interested in the annular structures of the inner disk, so we chose to ignore in-plane radiation transport. 
Nevertheless, to check for possible effects of in-plane radiation transport on the vortex dynamics, we repeated the $\alpha=10^{-5}$, $\beta=1$ model at 8~cps.
This time we included a flux-limited diffusion (FLD) approach \citep{levermore-pomraning-1981} similar to \citet{ziampras-etal-2020a}, but by parametrizing the diffusion coefficient $D_\mathrm{rad}$ following Eqs.~(12)--(14) of \citet{flock-etal-2017}:
\begin{equation}
	\label{eq:cooling-timescale-diffusion}
	t_\mathrm{cool} \approx \frac{l^2_\mathrm{thin}}{3D_\mathrm{rad}} + \frac{H^2}{D_\mathrm{rad}} = \beta\Omega_\mathrm{K}^{-1} \Rightarrow D_\mathrm{rad} = \frac{\Omega_\mathrm{K}}{\beta}\left(\frac{l^2_\mathrm{thin}}{3} + H^2\right),
\end{equation}
where $l_\mathrm{thin}$ is the photon mean free path. We found that including FLD slightly changes the radial surface density structure in the inner disk as predicted by \citet{miranda-rafikov-2020b} and reduces the vortex lifetime from 1200 to 900 orbits. Studying the effect of in-plane radiation transport in more detail requires further investigation.
\subsection{About the assumption of a 2D disk}
One of the main limitations of our models is the 2D assumption which was done due to runtime constraints in our rather wide exploration of the parameter space.
It is entirely possible that various 3D effects can result in quantitative differences in vortex properties.
3D vortices can be susceptible to the elliptical instability \citep{lesur_stability_2009} which would lower their lifetime. On the other hand, the vertical modes of the SBI could provide an additional channel to sustain the vortices, and vertical gas circulation due to the VSI might interfere with vortex growth and decay \citep{flock_gas_2020}.

To estimate the impact of including full-3D effects, we ran one 3D simulation with \texttt{FARGO3D} \citep{benitez-llambay_fargo3d:_2016} using a setup analogous to our 2D setup.
We chose $\beta=2\pi$, $\alpha=0$ and a resolution of 8~cps in all three directions.
The simulation assumed symmetry about the midplane and covered four scale heights in the vertical direction.
Similar to our 2D models, a large vortex formed at the outer gap edge and lived for 7000~orbits.
This illustrates that while there are differences, large vortices can survive in 3D disk simulations for a long time, even longer than in 2D for our example.
We limited the 3D runs to this one test because its runtime at 8~cps resolution was close to 4 months with the simulation performed on 4 NVIDIA K80 GPUs.
\subsection{On the observability of vortices at large radii}

Sect.~\ref{sec:lifetime-beta} illustrates that vortex lifetime is affected by the thermal cooling timescale $\beta$. The latter is expected to vary with radius in a disk, with values of 1--10 at 5\,au, 0.1 at $\sim$10\,au, and below 0.1--0.01 at $\sim$50\,au \citep{ziampras-etal-2020b}. 
Thus, we expect vortices to be in the short-lived regime close to the star and in the long-lived regime far from the star.
From Fig.~\ref{fig:masterplot} we can estimate the lifetime of vortices in disks with $\alpha \leq 10^{-4}$ to be between 500 and 3000~orbits for $\beta \geqslant 1$ and between 1000 and 15\,000~orbits for $\beta < 1$ for $\alpha \leq 10^{-4}$.
Assuming a solar-mass star, this yields estimated lifetimes for a planet-induced vortex between 6--30\,kyr at 5\,au, 175--700\,kyr at 50\,au, and 1--15\,Myr at 100\,au.
On the basis of a simple lifetime-centered argument, our results, therefore, suggest that planet-induced vortices are more likely to be observed at larger radii.

It should be noted, that planet growth timescales of 100 and 1000 planetary orbits are at the very low end of the spectrum of physically expected planet growth times.
\citet{hammer_slowly-growing_2017} provided estimates for more realistic planet growth-times of several thousand up to tens of thousands of orbits.
It remains to be seen, whether the effects observed in this study still appear for longer, more realistic, planet growth timescales.
However, simulating the disks at the required resolution of at least 16~cps for longer planet growth times along with the additional vortex evolution time is still computationally expensive.

\subsection{On using the lifetime of vortices in simulations to explain observations}
\label{sec:vortex-lifetime}

In the suite of simulations we carried out, the lifetime of vortices in models with identical physical parameters varies significantly with resolution.
This was the case for low values of the viscous $\alpha$ parameter ($\alpha  = 10^{-5}, 10^{-6}$).
We argued that the numerical viscosity of our simulation codes is
comparable to $\alpha_\text{num} \lesssim 10^{-5}$.
This suggests that simulations with a prescribed viscosity of the order of the numerical viscosity cannot be
used as a controlled numerical experiment, at least as far as the occurrence and persistence of vortices is concerned.
For prescribed viscosities well above the estimated numerical viscosity ($\alpha = 10^{-4}$, $10^{-3}$ in our case),
the consistency of vortex lifetimes between the two codes and numerical choices supports the idea that the numerical experiment is indeed a controlled one.

Recent observations of molecular line broadening \citep[e.g.,][]{flaherty_turbulence_2018}, numerical studies of VSI turbulence \citep[e.g.,][]{flock-etal-2017} and planet--disk interaction \citep[e.g.,][]{zhang-etal-2018} point to low $\alpha$ values. 
The requirement of a numerical viscosity lower than the physical viscosity necessitates high resolution, which poses a challenge for simulations of vortices in protoplanetary disks.

\subsection{Resolution and numerical viscosity}
\label{sec:numerical-viscosity}

Vortex evolution in ``inviscid'' disks is often studied using very high-resolution grids to minimize the effects of numerical viscosity \citep{li_potential_2005,paardekooper_vortex_2010,lin_effect_2011,zhu_gas_2016,hammer_slowly-growing_2017,mcnally_migrating_2019,hammer_which_2021, fung_cooling-induced_2021}.
While the resolution of 8 and 16 cells per scale height is likely enough to resolve planet-generated features such as the gap shape and spiral arms (see Appendix~\ref{sec:appendix-grid-resolution}), the numerical viscosity also needs to be low enough not to interfere with vortex decay.

An estimation of the numerical viscosity, valid for first-order schemes, is $\nu_\text{num} \sim \frac{\Delta x^2}{\Delta t}$, with a representative cell size $\Delta x$ and the timestep $\Delta t$.
For our choices of parameters and assuming $\Delta t \approx \frac{\Delta x}{c_s}$ this corresponds to $\alpha_\text{num} \sim 10^{-2}$--$10^{-1}$.
Clearly, we see substantial changes in dynamics down to much lower values of the prescribed $\alpha$.
Because we employ a higher-order scheme, this simple estimate is not applicable.
To our knowledge, there exists no formula to estimate the numerical viscosity for the higher-order schemes employed in this study, so we attempt to estimate it by comparing the results of our simulations at different values of $\alpha$.

In general, we observe a similar behavior between models with $\alpha=10^{-5}$ and $10^{-6}$, both in terms of the behavior of vortices during their lifetime (size, mass, migration patterns) as well as the overall lifetime itself (see Fig.~\ref{fig:masterplot}). This is also true across both codes that we used in this study, with the exception of the 8~cps models for $\beta=100$. 
We attribute the similarity to the numerical diffusion inherent in the different advection schemes of the two codes and expect that this translates to an effective $\alpha_\mathrm{num}$ between $10^{-6}$ and $10^{-5}$ for our given choices of grid resolution. This implies that our experiments with $\alpha=10^{-6}$ are most likely not controlled ones, and for this reason, we typically group models with $\alpha\leq10^{-5}$ together.

Nevertheless, we still observe a different behavior for some models with $\alpha=10^{-6}$ when comparing them to those with $\alpha=10^{-5}$, such as the migration of the long-lived models presented in Sect.~\ref{sec:long-lived} (see the different tracks of $r_\mathrm{vort}(t)$ in Fig.~\ref{fig:ts_outliers}), most of which have a 16~cps resolution. This hints at a lower numerical diffusion for 16~cps of $\alpha_\mathrm{num}\sim10^{-6}$.
Because the numerical viscosity in the 8~cps models might interfere with the prescribed $\alpha\leq10^{-5}$, our 8~cps simulations might not be as trustworthy as our higher-resolution 16~cps, $\alpha\leq10^{-5}$ runs.

\subsection{On the different numerics of the two codes}
\label{sec:fargo-vs-pluto-differences}

We used two codes (\texttt{PLUTO} and \texttt{FargoCPT}) with fundamentally different numerical properties. The fact that the two codes agree in terms of results (see the orange lines in Fig.~\ref{fig:ts_outliers} for one striking example) is reassuring, 
but it is worth discussing their differences nonetheless. 

\texttt{FargoCPT} requires an artificial viscosity prescription to stabilize the upwind method near regions of strong compression such as shocks.
This provides additional dissipation which could affect the evolution of vortices whenever they interact with the spiral shocks induced by the planet.
With the exception of the 8~cps models for $\beta=100$, we found no significant differences in vortex lifetimes between the two codes.
The one case for which the codes disagreed might be a result of insufficient resolution because the differences disappear for 16~cps.

On the other hand, \texttt{PLUTO}'s strictly energy-conserving nature means that the evolved quantity in the energy equation is the sum of kinetic and thermal energy.
Since kinetic energy dominates over thermal in typical Keplerian flows (for our setup, $E_\mathrm{kin}/E_\mathrm{th}\approx \frac{\gamma-1}{2h^2} = 80$),
numerical errors in the calculation of total energy could affect the thermal energy budget of the disk due to subtractive cancellation error.
In order to check this effect, we reran our fiducial model using the \texttt{ENTROPY\textunderscore SWITCH} option of \texttt{PLUTO},
which ensures entropy conservation outside of the vicinity of shocks
(which by definition do not conserve entropy, but are captured accurately by the Riemann solver).
We found that this did not affect the life track of the generated vortex.

Finally, we also reran the fiducial model with \texttt{PLUTO} using a 3rd-order solver and parabolic reconstruction instead of the standard 2nd-order solver and linear reconstruction setup.
We found no differences in vortex evolution or lifetime.

On the basis of our tests and the agreement of the codes for high resolution, we conclude that the vortex dynamics and effects we observed in our simulations are not numerical artifacts but that they are indeed physical.

\section{Summary}
\label{sec:summary}

We studied vortices created by planets in protoplanetary disks using two-dimensional viscous hydrodynamics simulations.
The equation of state was assumed to follow an ideal gas, turbulence was included following the $\alpha$ parametrization, and thermal processes were considered by prescribing a thermal relaxation timescale using the $\beta$ formalism.
A focus was brought to vortices exterior to the gap opened by the planet.
In order to verify our results, the simulations were carried out with both the \texttt{FARGO} and \texttt{PLUTO} codes which use different numerical schemes. 
The planet was treated as a non-accreting point mass with a smoothed gravitational potential and kept on a fixed circular orbit.
Properties of vortices were automatically extracted using our newly-developed \texttt{Vortector} Python tool, which identifies and characterizes vortices.
Vortex identification was performed by looking for elliptical shapes in iso-vortensity lines in the $r$-$\phi$ plane, and characterization was performed by fitting a 2D Gaussian to the vortensity and surface density.

Vortices formed during the gap opening process as the embedded Jupiter-mass planet was introduced into the simulation.
At the outer gap edge, multiple small vortices formed that usually merged into a single large vortex that lived, depending on parameters,
between 200 and several thousand orbits.
These vortices had a full width at half maximum (as determined by the fitted 2D Gaussian) of up to $0.4\,r_\mathrm{p}$ (several au for a planet at $r_\mathrm{p}=5.2$\,au).
The mass enclosed in this vortex area was up to one planetary mass (one Jupiter-mass in our models) for our choice of disk mass.

Vortex lifetime depended on the thermal relaxation timescale such that vortices lived shortest for intermediate cooling times ($\beta = 1$), a result also found by \citet{fung_cooling-induced_2021}.
We found two regimes for the vortices' lifetimes.
A short-lived regime, with vortex lifetimes of up to 3000 orbits, was observed for slowly-cooling disks ($\beta \geq 1$), in which the vortices decayed faster than expected from viscous dissipation alone.
In the long-lived regime, which was observed for fast cooling ($\beta \ll 1$) with the isothermal assumption as an extreme, vortices lived for a much longer time and did not decay rapidly. 
Vortex lifetimes were considerably longer in this regime, with a lower bound on the maximum lifetime being 15\,000 orbits (the model was terminated while the vortex was still alive due to runtime constraints).
From our analysis, we suspect that the long lifetime for small $\beta$ is connected to the interaction of the vortex with the spiral arms, which are a source of vorticity. Details are left to future studies.

Additionally, including the disk's self-gravity in our models, with a Toomre parameter $Q \approx 25$, usually shortened the lifetime of vortices and stopped the small initial vortices from merging into one large vortex.
Typically, in our models, two smaller vortices remained after the initial gap opening process, which then decayed faster compared to those in models where disk self-gravity was not accounted for.
This finding, that self-gravity is detrimental to vortex survival, is in line with previous studies \citep{lovelace_rossby_2013,zhu_gas_2016,regaly_vortex_2017,pierens_evolution_2018}.

Outward migration of the vortex was observed in some of the models with $\beta \ll 1$ and $\beta \gg 1$.
In those cases, a second density (and thus pressure) bump formed outside of the vortex location, towards which the vortex then migrated \citep{paardekooper_vortex_2010}. 
In some $\beta=0.01$ models, a small, short-lived, ``secondary'' vortex formed between the planet gap and the ``primary'' vortex.

Concerning the dependence of vortex lifetime on viscosity, we found the expected behavior that this lifetime was shorter for higher viscosity \citep{godon_vortices_1999,val-borro_vortex_2007,ataiee_asymmetric_2013,fu_long-term_2014,regaly_interpreting_2017}.
For the highest viscosity of $\alpha=10^{-3}$, practically no vortices were observed.
For $\alpha = 10^{-5}$ and $10^{-6}$ we found nearly identical results, suggesting that the numerical viscosity in our models with a resolution of 8 and 16 cells per scale height was of the order of $\alpha_{8\text{cps}} \lesssim 10^{-5}$ and $\alpha_\text{16\text{cps}} \approx 10^{-6}$.

Allowing the planet to grow over a longer time, 1000 instead of 100 orbits, led to longer vortex lifetimes in all the cases we tested.
This disagrees with the findings of \citet{hammer_slowly-growing_2017}, who found reduced vortex lifetimes for longer planet-growth times.
In our models, vortices took longer to form in the case of the slower-growing planet.
During their decay, however, their evolution was very similar, independent of planet introduction time (see Fig.~\ref{fig:ts_rampup}), which in total increased their lifetime.
The fact that vortex lifetime increased for longer planet-growth timescales can be an indication that the effects presented in this study, including the long-lived vortex regime, are also applicable to longer, and arguably more realistic, planet-growth timescales of around 10\,000 orbits.

Estimating vortex lifetime from our results, vortices are expected to live much longer at larger distances away from their host star.
The increase in expected lifetime is firstly due to the longer orbital period at large radii, but also because the expected $\beta$ values --- the thermal relaxation timescale compared to the orbital timescale -- are much lower and vortices then likely belong to the
long-lived regime (see Sect.~\ref{sec:long-lived}).
From order-of-magnitude calculations, we find that large planet-induced vortices exterior to the planet at 50--100\,au might live for up to several Myr for low-viscosity disks ($\alpha \lesssim 10^{-4}$).
Considering the sensitivity of instruments like ALMA at these distances from the star, this suggests that these vortices should be observable more easily than planet-induced vortices at smaller radii.

\begin{acknowledgements}
	TR and WK acknowledge funding from the Deutsche Forschungsgemeinschaft (DFG) research group FOR 2634 ''Planet Formation Witnesses and Probes: Transition Disks''
	under grant DU 414/22-1 and KL 650/29-1, 650/30-1.
	WB, WK, and AZ acknowledge support by the DFG-ANR supported GEPARD project (ANR-18-CE92-0044 DFG: KL 650/31-1). 
	The authors acknowledge support by the High Performance and Cloud Computing Group at the Zentrum
	f\"ur Datenverarbeitung of the University of T\"ubingen, the state of Baden-W\"urttemberg
	through bwHPC and the German Research Foundation (DFG) through grant INST\,37/935-1\,FUGG.
	Plots in this paper were made with the Python library \texttt{matplotlib} \citep{hunter-2007}.
  \end{acknowledgements}

\bibliographystyle{aa}
\bibliography{paper}

\appendix

\section{Grid resolution and numerical convergence}
\label{sec:appendix-grid-resolution}

A grid resolution of 8 cells per scale height (cps) is often adopted in models of planet--disk interaction in literature.
While it is widely agreed upon as sufficient, we test this statement by performing a series of test simulations with both \texttt{PLUTO} and \texttt{FargoCPT} using the same physical parameters as our locally isothermal models ($\alpha=10^{-5}$),
but using varying grid resolutions with 1, 2, 4, 8 and 16~cps in both directions (always maintaining square cells).

We found that we achieve numerical convergence on large-scale features such as the gap width and pressure bumps for a resolution of 4 cps. Convergence on more numerically-sensitive features such as gap depth and vortex formation is reached for a resolution of 8~cps,
with 16~cps affecting the picture relatively weakly.
This was observed across both codes,
with the two showing very good agreement with each other both in terms of the resolution at which different features converge and the physical properties of said features across codes.

\begin{figure}[ht]
	\centering
	\includegraphics[width=.5\textwidth]{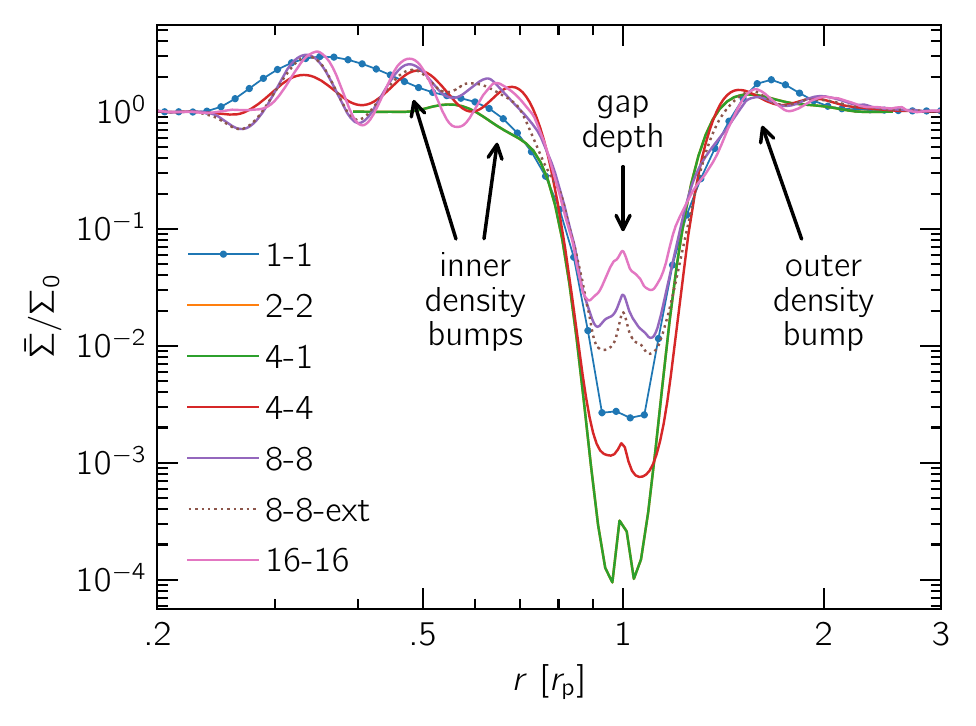}
	\caption{Results for our resolution study using \texttt{FargoCPT}. The overall shape of the gap is resolved with around 4 cps, while it takes 8 cps to properly resolve the gap depth and the contrast of most pressure bumps far from the gap. We are interested in the region between 0.5--2.0\,$r_\mathrm{p}$. Extending the outer boundary to $r=10\,r_\mathrm{p}$ in the ``8-8-ext'' model practically made no difference. It should be noted that the $16\times16$~cps model develops some small-scale vortices in the inner disk, which causes these differences around 0.7\,$r_\mathrm{p}$. Interestingly, a model that resolves the radial and azimuthal directions with 4 and 1 cps, respectively, captures these radial features almost as well as one with 4 cps in both directions.}
	\label{fig:resolution-sigma}
\end{figure}

\section{The Vortector}\label{sec:vortector}

A major task in this study was the identification and characterization of vortices in simulation data.
For this purpose, we developed a Python package, the \texttt{Vortector}, that automates the process for relatively generic 2D hydrodynamics planet--disk simulations.

The \texttt{Vortector} package lets one visualize the vortex detection results (an example is shown in Fig.~\ref{fig:vortector_overview}), as well as includes information about the location, extent, and mass of a vortex along with various statistics related to the contour.

The package is publicly available on GitHub\footnote{\url{https://github.com/rometsch/vortector}}. 
We hope to make the detection and characterization of vortices in simulation data easier for other members of the community and facilitate quantitative comparison of vortices between studies by providing a common detection pipeline.

To search for possible vortex candidates, a simple search for the location of minimum vortensity is sometimes enough to find the location of a vortex.
Then, the value of the vorticity $\omega = (\nabla \times \vec{u}) \cdot \hat{z}$ can be used to learn how strongly the vortex rotates and the local surface density can be used as an indication for the mass enclosed in the vortex.
This method, however, fails for many simulations,
e.g., when the vortensity in a tiny region close to a spiral arm of the planet is lower than inside a vortex candidate,
or when the gap region intrudes into the outer disk, which can induce strong anticyclonic motion at the outer gap edge.

To get around these issues, the \texttt{Vortector} uses the geometrical shape of vortices as they appear in a face-on image of the disk.
Looking down on the surface of a disk, vortices appear as crescent-shaped objects.
In the $r$-$\phi$ plane, which is more suitable for this task,
large vortices appear as elliptical objects \citep[see also Fig. 1 of][]{lesur_stability_2009}.
In fact, contour lines of the vortensity closely resemble ellipses in the $r$-$\phi$ plane.
We can therefore identify vortices in a disk by finding closed contour lines that closely resemble ellipses.
To solve this task programmatically, we can make use of the computer vision library \texttt{OpenCV} \citep{opencv_library}.

Our strategy to extract vortex candidates from a simulation snapshot can be then subdivided into three tasks:

\begin{enumerate}
	\item Extract contour lines in the $r$-$\phi$ map of the vortensity, 
	\item identify nearly elliptical contours as vortex candidates, and
	\item fit 2D Gaussians to $\varpi$ and $\Sigma$ for characterization.
\end{enumerate}

\begin{figure}[t]
	\centering
	\includegraphics[width=0.5\textwidth]{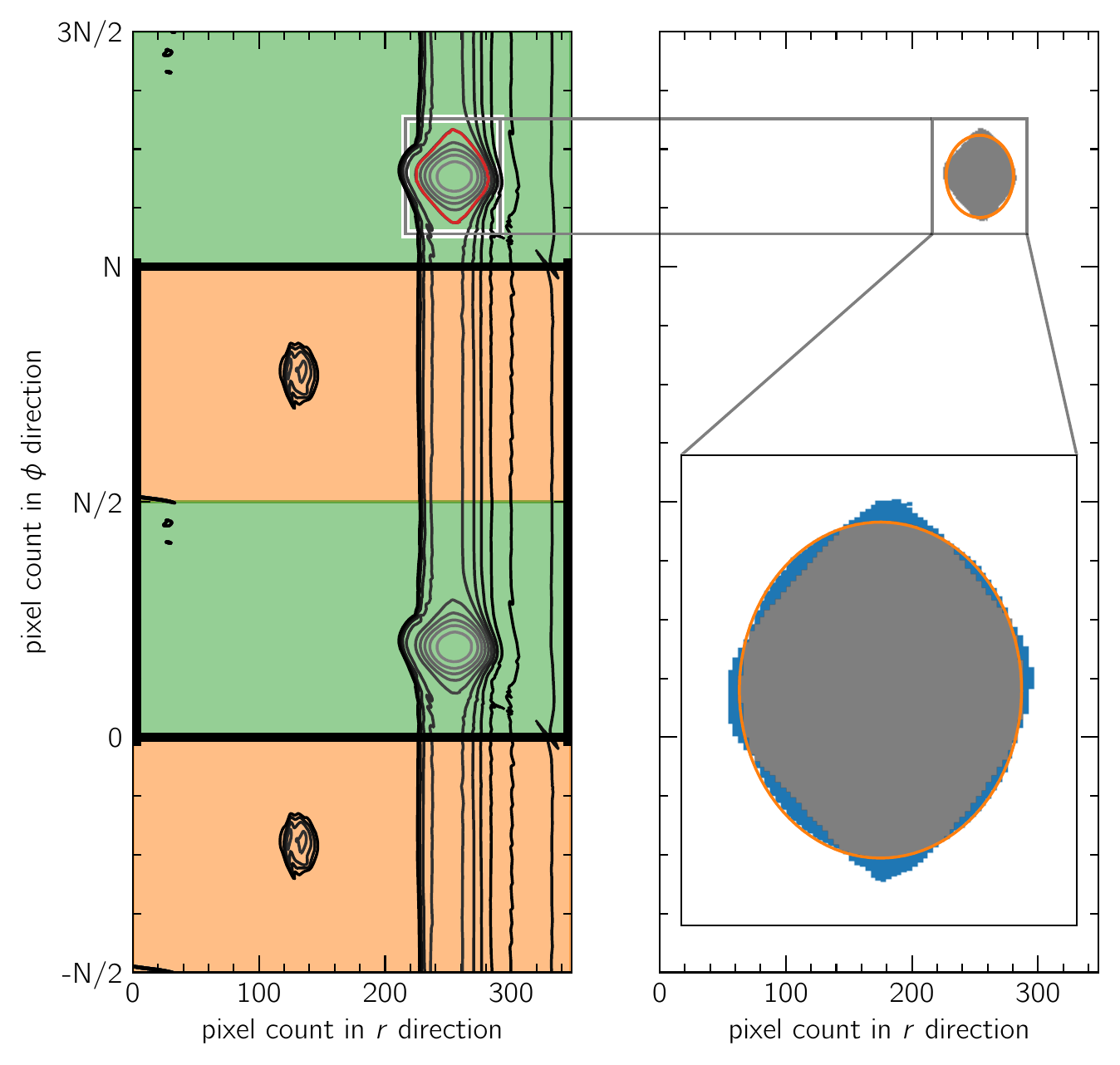}
	\caption{Periodically continued iso-vortensity line image (left) used for extracting contours for vortex candidates and an example contour illustrating the ellipse fit (right).
		The snapshot shown is at time $t=7150\,\mathrm{orbits}$ of the model with the ``secondary'' vortex that was discussed in Sect.~\ref{sec:long-lived}.
		The left panel shows how the data array is mirrored in order to allow the detection of vortices that overlap with the periodic boundary.
		Areas with the same color are copies of one another.
		The red line indicates the outline of the grey area in the right panel.
		The original size is marked by the black rectangle ranging from $0$ to $N$ on the vertical axis.
		The area shaded in blue in the right panel illustrates the definition of the deviation from the ellipse that is used to select the vortex candidates from the closed contours.
		For the example shown, the ratio of difference area (blue) to the total contour area is 0.122, which is below the 0.15 threshold.
		\label{fig:contour_fits}}
\end{figure}

\subsection*{The algorithm step by step}

This section describes the vortex detection process using the model presented in Sect.~\ref{sec:long-lived} (\texttt{FargoCPT}, $\alpha=10^{-5}$, $\beta=0.01$, 16~cps) which shows the emergence of a secondary vortex.
The data used for this analysis corresponds to a time $t = 7150\,\mathrm{orbits}$.

Before the analysis is performed on $\varpi$, the map is periodically extended in the $\phi$ direction in order to be able to identify vortices that intersect the periodic azimuthal boundary.
The resulting image is shown in the left panel of Fig.~\ref{fig:contour_fits}.
There, the original domain is indicated by the thick solid rectangle which spans the azimuthal range from 0 to $N$, where $N$ is the image size in pixels in the azimuthal direction.
The top and bottom areas of the domain (orange and green) are repeated at the lower and upper boundaries, respectively.

\subsubsection*{Task 1: contour lines}

Contour lines are extracted for a range of $\varpi$ values ranging from 0 to 1 in increments of 0.05.
For each value $\varpi_\mathrm{crit}$, a binary image is produced by setting each cell with $\varpi \geq \varpi_\mathrm{crit}$ to 1 and 0 otherwise.
The binary image is then analyzed using \texttt{findContours} from \texttt{OpenCV}.
Only closed contours are retained.
This step usually results in up to a few thousand contours, depending on the dynamical state of the disk and the choice of increments in $\varpi$.

\subsubsection*{Task 2: Find closed contours resembling ellipses}

Next, the \texttt{fitEllipse} function from \texttt{OpenCV} is used to fit an ellipse to each closed contour.
One example of this is shown in the right panel of Fig.~\ref{fig:contour_fits}, where the ellipse is visible as an orange line in the zoom-in.

The difference in area between the contour and fit is used as a measure of deviation.
The deviation from an ellipse is then defined as the ratio of this difference and the area enclosed by the contour.
We only keep contours for which the deviation is smaller than 0.15.
The example contour in Fig.~\ref{fig:contour_fits} has a deviation of 0.122.

\begin{figure*}[t]
	\centering
	\includegraphics[width=\textwidth]{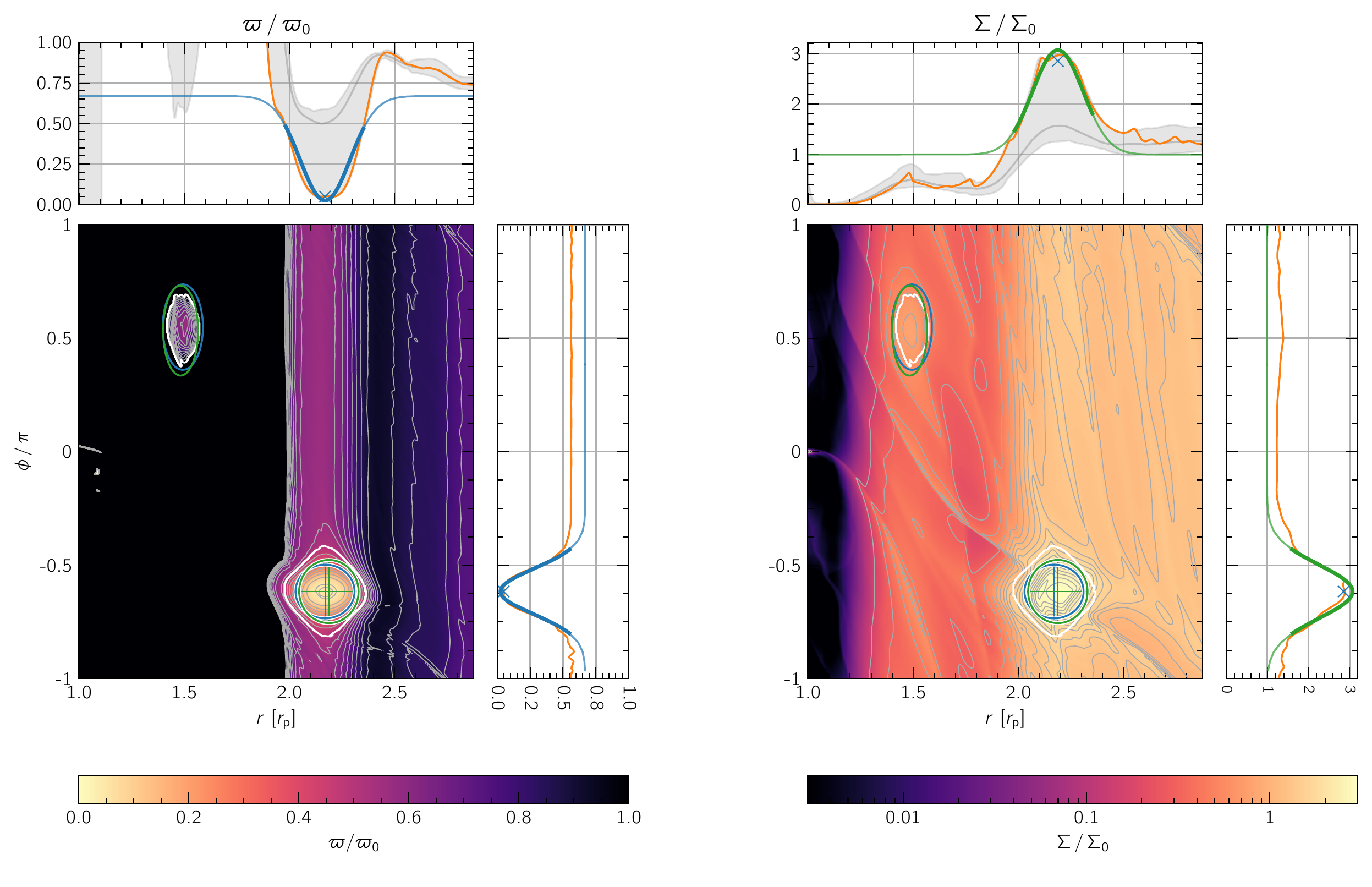}
	\caption{Overview of the results produced by the \texttt{Vortector} package for a model showing a secondary vortex discussed in Sect.~\ref{sec:long-lived} at $t = 7150\,\mathrm{orbits}$, with a 2D map of the vortensity on the left and surface density on the right.
		All detected vortex candidates are indicated in the 2D plots.
		The extracted contour (shown in Fig.~\ref{fig:contour_fits}) is marked with a white line, the ellipse of the vortensity fit is shown in blue and the ellipse of the surface density fit is shown in green.
		Note that these ellipses are defined by $\sigma_r$ and $\sigma_\phi$ from the fit of Eq.~\eqref{eqn:double_gauss} and are different from the ellipse used to fit the contour. 
		The ellipses of the most massive vortex include a crosshair indicating the center of the fit.
		Each 2D plot is accompanied by 1D plots of slices through the main vortex.
		They also show the values of the respective Gaussian fit in blue for vortensity and green for surface density.
		In this figure, the planet is located at $r=1$ and $\phi = 0$.
		\label{fig:vortector_overview}}
\end{figure*}

Finally, all the contours that are contained within the largest contour that satisfies this criterion are discarded, which leaves the example red contour in Fig.~\ref{fig:contour_fits} as the selected vortex candidate (see also the white contour line in Fig.~\ref{fig:vortector_overview}).

We only retain contours that enclose at least two other contours.
With this restriction, we make sure that $\varpi/\varpi_0$ changes by a value of 0.1 from the outside to the inside of the vortex candidate.
This has proven to be useful to filter out small fluctuations in the disk that otherwise appear as small transient vortices.

At this point, it becomes clear that the extent of the vortex and derived quantities such as the mass contained within are influenced by the choice of the levels used to produce the contour lines and the choice of the maximum relative ellipse deviation.
The properties of the contour give an order-of-magnitude estimate nonetheless.

\subsubsection*{Task 3: Fit a 2D Gaussian}

To remove the influence of the threshold parameters in the detection of the contour, a process that does not depend on our parameter choices but on the underlying data is needed.

Upon inspection of the curves of vortensity and surface density along a cut through the vortex, either radial or azimuthal, it becomes clear that these lines resemble Gaussian functions (see curves in Fig.~\ref{fig:vortector_overview} around the 2D maps)
\begin{align}\label{eq:gaussian_2D}
	f(r, \phi) = c + a \exp\left( - \frac{(r - r_0)^2}{2 \sigma_r^2} \right) \exp\left( - \frac{(\phi - \phi_0)^2}{2 \sigma_\phi^2} \right)\,.
\end{align}

Here, $\sigma_r$ and $\sigma_\phi$ provide a measure for the vortex size and can even be used to give a definition of the vortex region that does not depend on additional parameters.
In combination with the center coordinates $r_0$ and $\phi_0$, $\sigma_r$ and $\sigma_\phi$ can be used to define the vortex as the disk material contained within the ellipse given by
\begin{align}\label{eqn:double_gauss}
	\left( \frac{r - r_0}{h_r} \right)^2 + \left( \frac{\phi - \phi_0}{h_\phi} \right)^2 = 1\,,
\end{align}
where $h_r = \sqrt{2\ln(2)}\, \sigma_r$ and $h_\phi = \sqrt{2\ln(2)}\, \sigma_\phi$ denote the half width at half maximum of the 2D Gaussian function defined in Eq.~\eqref{eq:gaussian_2D}.
We usually use the values obtained from the surface density fit because these are less time-sensitive compared to the vortensity fit and because the shape of $\Sigma$ curves more closely resemble Gaussians (see Fig.~\ref{fig:vortector_overview}).

\section{Data table}

The lifetimes and parameters of all models mentioned in Sects.~\ref{sec:results} and \ref{sec:long-lived} are listed in Table~\ref{tab:lifetimes}.

\begin{table*}[ht]
	\small
	\begin{center}
		\caption{Lifetimes of vortices in the simulations.}
		\label{tab:lifetimes}
		\begin{tabular}{|cccccccc|c|cccccccc|}
			code & $\beta$ & $\alpha$ & cps & $\tau_\mathrm{intro}$\tablefootmark{a} & special\tablefootmark{b} & in Fig.~\ref{fig:masterplot} & $T_\mathrm{vort}$\tablefootmark{c} & & code & $\beta$ & $\alpha$ & cps & $\tau_\mathrm{intro}$ \tablefootmark{a} & special\tablefootmark{b} &  in Fig.~\ref{fig:masterplot} & $T_\mathrm{vort}$\tablefootmark{c} \\
			\hline
			\input{datatable_double_data.tex}
		\end{tabular}
		\tablefoot{
			Models with a resolution of 8, 16, and 32 cps have $528\times1024$, $1056\times2048$, and $2112\times4096$ cells, respectively.
			\tablefoottext{a}{Planet introduction time. 100 orbits if empty.}
			\tablefoottext{b}{Special propertiy of the model. ``sg'' if self-gravity is included. ``vort rem'' and ``art vort'' refer to the models discussed in Sect.~\ref{sec:vortex-conditions} with the removed vortex and the artificial vortex, respectively.}
			\tablefoottext{c}{Lifetime of the vortex in planetary orbits. ``>'' indicates that the vortex still exists at the end of the simulation.}
		}
	\end{center}
\end{table*}

\end{document}

%% file: datatable_double_data.tex
fargo & 0.01 & $10^{-6}$ & 16 & 100 &  & x & >8784 & & fargo & 1 & $10^{-5}$ & 8 & 1000 &  &  & 1279\\
pluto & 0.01 & $10^{-6}$ & 16 & $\dots$ &  & x & 9548 & & fargo & 1 & $10^{-5}$ & 8 &  &  & x & 1179\\
fargo & 0.01 & $10^{-6}$ & 8 &  &  & x & 1679 & & fargo & 1 & $10^{-5}$ & 8 &  & sg & x & 289\\
pluto & 0.01 & $10^{-6}$ & 8 &  &  & x & 2013 & & fargo & 1 & $10^{-5}$ & 8 & 1000 & sg &  & 509\\
fargo & 0.01 & $10^{-5}$ & 32 &  &  &  & >739 & & pluto & 1 & $10^{-5}$ & 8 &  &  & x & 1136\\
fargo & 0.01 & $10^{-5}$ & 16 &  &  & x & 8774 & & pluto & 1 & $10^{-5}$ & 8 & 1000 &  &  & 956\\
fargo & 0.01 & $10^{-5}$ & 16 &  & sg &  & >4567 & & fargo & 1 & $10^{-4}$ & 16 &  &  & x & 709\\
pluto & 0.01 & $10^{-5}$ & 16 &  &  & x & >15100 & & pluto & 1 & $10^{-4}$ & 16 &  &  & x & 757\\
pluto & 0.01 & $10^{-5}$ & 16 &  &  &  & 5762 & & fargo & 1 & $10^{-4}$ & 8 &  &  & x & 659\\
pluto & 0.01 & $10^{-5}$ & 16 &  & vort rem &  & 1843 & & pluto & 1 & $10^{-4}$ & 8 &  &  & x & 607\\
fargo & 0.01 & $10^{-5}$ & 8 & 1000 &  &  & 3897 & & fargo & 1 & $10^{-3}$ & 16 &  &  & x & 0\\
fargo & 0.01 & $10^{-5}$ & 8 &  &  & x & 2288 & & pluto & 1 & $10^{-3}$ & 16 &  &  & x & 49\\
fargo & 0.01 & $10^{-5}$ & 8 &  &  &  & 7845 & & fargo & 1 & $10^{-3}$ & 8 &  &  & x & 59\\
fargo & 0.01 & $10^{-5}$ & 8 &  & sg & x & 1299 & & pluto & 1 & $10^{-3}$ & 8 &  &  & x & 0\\
fargo & 0.01 & $10^{-5}$ & 8 & 1000 & sg &  & 1269 & & fargo & 100 & $10^{-6}$ & 16 &  &  & x & 1729\\
pluto & 0.01 & $10^{-5}$ & 8 &  &  & x & 2611 & & pluto & 100 & $10^{-6}$ & 16 &  &  & x & 2441\\
pluto & 0.01 & $10^{-5}$ & 8 & 1000 &  &  & 2691 & & fargo & 100 & $10^{-6}$ & 8 &  &  & x & 3108\\
pluto & 0.01 & $10^{-5}$ & 8 &  &  &  & 10638 & & pluto & 100 & $10^{-6}$ & 8 &  &  & x & 946\\
fargo & 0.01 & $10^{-4}$ & 16 &  &  & x & 1129 & & fargo & 100 & $10^{-5}$ & 16 &  &  & x & 1888\\
pluto & 0.01 & $10^{-4}$ & 16 &  &  & x & 1335 & & pluto & 100 & $10^{-5}$ & 16 &  &  & x & 1943\\
fargo & 0.01 & $10^{-4}$ & 8 &  &  & x & 979 & & fargo & 100 & $10^{-5}$ & 8 & 1000 &  &  & 3048\\
pluto & 0.01 & $10^{-4}$ & 8 &  &  & x & 887 & & fargo & 100 & $10^{-5}$ & 8 &  &  & x & 2708\\
fargo & 0.01 & $10^{-3}$ & 16 &  &  & x & 79 & & fargo & 100 & $10^{-5}$ & 8 &  & sg & x & 589\\
pluto & 0.01 & $10^{-3}$ & 16 &  &  & x & 0 & & fargo & 100 & $10^{-5}$ & 8 & 1000 & sg &  & 499\\
fargo & 0.01 & $10^{-3}$ & 8 &  &  & x & 69 & & pluto & 100 & $10^{-5}$ & 8 &  &  & x & 857\\
pluto & 0.01 & $10^{-3}$ & 8 &  &  & x & 29 & & pluto & 100 & $10^{-5}$ & 8 & 1000 &  &  & 0\\
fargo & 1 & $10^{-6}$ & 16 &  &  & x & 1309 & & fargo & 100 & $10^{-4}$ & 16 &  &  & x & 699\\
pluto & 1 & $10^{-6}$ & 16 &  &  & x & 1694 & & pluto & 100 & $10^{-4}$ & 16 &  &  & x & 797\\
fargo & 1 & $10^{-6}$ & 8 &  &  & x & 1189 & & fargo & 100 & $10^{-4}$ & 8 &  &  & x & 569\\
pluto & 1 & $10^{-6}$ & 8 &  &  & x & 1106 & & pluto & 100 & $10^{-4}$ & 8 &  &  & x & 478\\
fargo & 1 & $10^{-5}$ & 16 &  &  & x & 1359 & & fargo & 100 & $10^{-3}$ & 16 &  &  & x & 0\\
fargo & 1 & $10^{-5}$ & 16 &  & sg &  & 649 & & pluto & 100 & $10^{-3}$ & 16 &  &  & x & 0\\
pluto & 1 & $10^{-5}$ & 16 &  &  & x & 1445 & & fargo & 100 & $10^{-3}$ & 8 &  &  & x & 0\\
pluto & 1 & $10^{-5}$ & 16 &  & art vort &  & 916 & & pluto & 100 & $10^{-3}$ & 8 &  &  & x & \ 0